\documentclass[aip,jcp,amsmath,amssymb,reprint,superscriptaddress]{revtex4-2}
\usepackage{graphicx}
\usepackage{epsf}
\usepackage{bm}
\usepackage{physics}
\usepackage{lipsum}
\usepackage[normalem]{ulem}
\usepackage{txfonts}
\usepackage{booktabs}
\usepackage{multirow}
\usepackage[
    colorlinks=true,
    linkcolor=cyan,
    citecolor=blue,
    urlcolor=blue
]{hyperref}

\newcommand{\nocontentsline}[3]{}
\newcommand{\tocless}[2]{\bgroup\let\addcontentsline=\nocontentsline#1{#2}\egroup}

\graphicspath{{./}{./figs/}{../figs/}}

\begin{document}

\title{
Deep learning of committor and explainable artificial intelligence analysis for identifying reaction coordinates
}

\author{Toshifumi Mori}
\email{toshi\_mori@moleng.kyoto-u.ac.jp}
\affiliation{Department of Molecular Engineering, Graduate School of Engineering, Kyoto University, Kyoto 615-8510, Japan}
\affiliation{Institute for Materials Chemistry and
Engineering, Kyushu University, Kasuga, Fukuoka 816-8580, Japan}
\affiliation{Interdisciplinary Graduate School of Engineering Sciences,
Kyushu University, Kasuga, Fukuoka 816-8580, Japan}

\author{Kei-ichi Okazaki}
\email{keokazaki@ims.ac.jp}
\affiliation{Research Center for Computational Science, Institute for Molecular Science, Okazaki, Aichi 444-8585, Japan}
\affiliation{Graduate Institute for Advanced Studies, SOKENDAI, Okazaki,
Aichi 444-8585, Japan}

\author{Kang Kim}
\email{kk@cheng.es.osaka-u.ac.jp}
\affiliation{Division of Chemical Engineering, Department of Materials Engineering Science, Graduate School of Engineering Science, The University of Osaka, Toyonaka, Osaka 560-8531, Japan}

\author{Nobuyuki Matubayasi}
\email{nobuyuki@cheng.es.osaka-u.ac.jp}
\affiliation{Division of Chemical Engineering, Department of Materials Engineering Science, Graduate School of Engineering Science, The University of Osaka, Toyonaka, Osaka 560-8531, Japan}

\date{\today}

\begin{abstract}
In complex molecular systems, 
the reaction coordinate (RC) that characterizes transition pathways is
 essential to understand 
 underlying molecular mechanisms.
This review surveys a framework for identifying the RC by applying deep
 learning to the committor, which provides the most reliable measure of
 the progress along a transition path.
The inputs to the neural network are collective variables (CVs)
 expressed as functions of atomic coordinates of the system, and the
corresponding RC is predicted as the output by training the network on
 the committor as the learning target.
Because deep learning models typically operate in a black-box manner,
it is difficult to determine which input variables
govern the predictions.
The incorporation of eXplainable Artificial Intelligence (XAI)
 techniques enables quantitative assessment of 
the contributions of individual input variables to the predictions.
This approach allows the identification of CVs that play dominant roles
 and demonstrates that the committor distribution on the surface using
 important CVs
 is separated by well-defined boundaries.
The framework provides an explainable deep learning strategy for
 assigning a molecular mechanism from the RC and is applicable to
 a wide range of complex molecular systems.
\end{abstract}
\maketitle

\tableofcontents

\section{Introduction}

Understanding the mechanisms of transition processes in complex
molecular systems, as exemplified by protein conformational changes,
nucleation, and solvation effects, is essential for characterizing the
transition pathways that connect multiple stable states. 
However, the high dimensionality of these systems renders direct
visualization and interpretation inherently difficult. 
One widely used approach to address this challenge is the potential of
mean force (PMF).~\cite{zuckerman2010Statistical}

The PMF is referred to as the free-energy landscape, and more
specifically is defined as the free energy as a function of a
collective variable (CV) $x$, which is \textit{a priori} selected from the atomic
coordinate space $\mathbf{R}= (\mathbf{r}_1, \mathbf{r}_2, \cdots, \mathbf{r}_N)$
describing the positions of 
$N$ atoms: 
\begin{equation}
F(x)  = -k_\mathrm{B}T \ln \int  e^{-U(\mathbf{R})/k_\mathrm{B}T}
 \delta [x - x(\mathbf{R})] d\mathbf{R}, 
\label{eq:PMF}
\end{equation}
where $k_\mathrm{B}$, $T$, and $U(\mathbf{R})$ denote the Boltzmann
constant, temperature, and potential energy of the system.
Typical CVs are commonly defined in terms of interatomic distances, bond
angles, and dihedral angles that characterize molecular conformation
changes.
For example, the radial distribution function $g(r)$ with respect to the
interatomic distance $r$ can be interpreted as a PMF.
Calculating the PMF in molecular dynamics (MD) simulations of complex
molecular systems is a central challenge, and various sampling methods,
including umbrella sampling,~\cite{torrie1977Nonphysical} the replica
exchange method,~\cite{sugita1999Replicaexchange} and
metadynamics,~\cite{laio2002Escaping} have been developed to address the
high computational cost
associated with exhaustive 
configurational sampling.~\cite{chipot2007Free,
tuckerman2023Statistical, frenkel2023Understanding, chipot2023Free}

The term ``collective variable'' is often used interchangeable with ``order
parameter'', but the latter typically refer to a CV that distinguishes between
reactant and product states.
Furthermore, ``reaction coordinate'' can be regarded as an order parameter that characterizes the
dynamical progress from reactant to product states.~\cite{peters2017Reaction}
Specifically, within the framework of transition state theory, when a saddle point on
the PMF corresponds the transition state (TS), the selected CV serves as
the reaction coordinate (RC).~\cite{peters2017Reaction}
In this case, transitions from reactant to product states
necessarily proceed via barrier crossing along the RC.
However, identifying an appropriate CV that is truly relevant to the RC
among many candidate CVs is inherently nontrivial.
A critical issue is that the free-energy barrier on the PMF may
depend on the selection of CVs.~\cite{zuckerman2010Statistical,
frenkel2013Simulations, nakamura2024Derivation}
This highlights the necessity of rigorously examining whether a given
PMF can correctly capture the TS.

The committor analysis is a statistical approach for identifying the RC
from transition paths sampled in MD
simulations.~\cite{bolhuis2002TRANSITION, rohrdanz2013Discovering,
jungblut2016Pathways, banushkina2016Optimal, 
rogal2021Reaction, pietrucci2017Strategies}
Here, states A and B denote reactant and product states, respectively,
separated by the TS.
The committor $p_\mathrm{B}^*(\mathbf{R})$ is defined as the
probability of trajectories, initiated from an arbitrary configuration
$\mathbf{R}$ with velocities randomly drawn from the Maxwell--Boltzmann
distribution, that reaches state B prior to A.
If the initial configuration $\mathbf{R}$ locates at the TS, 
the probabilities of reaching A and B are equal, yielding $p_\mathrm{B}^*=1/2$.
In other words, the TS can be characterized as the set of configurations
exhibiting $p_\mathrm{B}^*=1/2$.
In practice, the committor analysis requires evaluating the distribution $p(p_\mathrm{B}^*)$
from an ensemble of configurations sampled in the vicinity of the saddle
point and assessing whether the selected CV is appropriate based on the presence of a sharp peak at 
$p_\mathrm{B}^*=1/2$
This committor distribution test has been widely applied to assess the
quality of selected CVs across a variety of complex molecular
systems.~\cite{du1998Transition, geissler1999Kinetic,
bolhuis2000Reaction, hagan2003Atomistic, hummer2004Transition,
pan2004Dynamics, ren2005Transition,
rhee2005OneDimensional, e2005Transition,
berezhkovskii2005Onedimensional, best2005Reaction, moroni2005Interplay,
maragliano2006String, 
peters2006Using, branduardi2007Free, quaytman2007Reaction,
antoniou2011Identification, ballard2012Mechanism, barnes2014Reaction,
pluharova2016Dependence, li2019Computing, brotzakis2019Approximating,
manuchehrfar2021Exacta, wu2022Rigorous,
wu2022Exact, hasyim2022Supervised, 
mochizuki2023Microscopic, li2025Enhanced, zhang2026Membrane}

The committor analysis is a reliable method; however, its
trial-and-error nature, which depends strongly on physical intuition,
limits its efficiency and motivates the development of more systematic
approaches.~\cite{peters2010Recent, li2014Recent, peters2016Reaction,
sittel2018Perspective}
Machine learning approaches based on committor analysis for identifying
RCs have attracted considerable attention and have led
to the development of various methodological frameworks. 
Ma and Dinner
introduced a genetic neural network method and applied it to committor
values evaluated for molecular configurations.~\cite{ma2005Automatic}
Peters \textit{et al.} developed a method based on maximum likelihood
estimation to identify the RC.~\cite{peters2006Obtaining, peters2007Extensions}
By combining aimless shooting, a variant of transition path sampling,
with a procedure that maximizes the log-likelihood, they introduced a
framework for quantifying the contributions of candidate CVs to the RC.
The aimless shooting yields a binary outcome for each trajectory
initiated from a given initial configuration, generating instantaneous committor values (or observed commitments)
$p_\mathrm{B}^*=0$ or 1, corresponding to states A and B, respectively.
The committor is modeled using a sigmoidal function,
$p_\mathrm{B}(q)=[1+\tanh(q)]/2$, 
and maximization of the long-likelihood based on these binary outcomes
determines the optimal RC.
Here, $q$ is ultimately expressed as a
linear combination of the candidate CVs associated with the initial
configurations. 
This likelihood-based optimization approach has been widely applied for
identifying RCs in a broad range of studies.~\cite{beckham2007SurfaceMediated, beckham2008Evidence,
peters2010TransitionState, 
vreede2010Predicting, lechner2010Nonlinear, pan2010Molecular,
beckham2011Optimizing, peters2012Inertial, juraszek2012Transition, xi2013Hopping,
jungblut2013Optimising, mullen2014Transmission, mullen2015Easy,
lupi2016Preordering, lupi2017Role, jung2017Transition, joswiak2018Ion,
diazleines2018Maximum, okazaki2019Mechanism, arjun2019Unbiased,
marriott2019Following, liang2020Identification,
rogers2020Breakage, schwierz2020Kinetic,
levintov2021Reaction, silveira2021Transitiona, li2022Optimizinga}
Other machine learning methodologies for identifying RCs that
characterize transition pathways have also been
developed.~\cite{schneider2017Stochastic, mcgibbon2017Identification, sultan2018Automated,
wehmeyer2018Timelagged, chen2018Molecular, ribeiro2018Reweighted, 
mardt2018VAMPnets, bittracher2018Datadriven, 
rogal2019NeuralNetworkBased, 
bonati2020DataDriven, 
bonati2021Deep, wang2021State, zhang2021Deep, hooft2021Discovering, magrino2022Critical, belkacemi2022Chasing,
baima2022Capabilities, sun2022Multitask, ketkaew2022DeepCV, 
lazzeri2023Molecular,
chen2023Discovering, singh2023Variationala, liang2023Probing, france-lanord2024DataDriven, 
zhang2024DescriptorFree, zhang2024Combining, herringer2024Permutationally,
frohlking2024Deep, kang2024Computing, mitchell2024Committor, 
megias2025Iterative, liu2025Memory, 
elangovan2025Datadriven, dietrich2025Reproducibility, pengmei2025Using,
prabhakar2025Discriminant, 
kresse2026Revealing, giuseppechen2026Following,
kang2026Committors, contrerasarredondo2026Learning}
Moreover, numerous review articles have appeared in recent
years.~\cite{wang2020Machine, sidky2020Machine, noe2020Machine,
chen2021Collective, 
fu2024Collective, mehdi2024Enhanced, 
singh2025Variational, gokdemir2025Machine, zhu2026Enhanced}

\begin{figure*}[t]
\centering
\includegraphics[width=\textwidth]{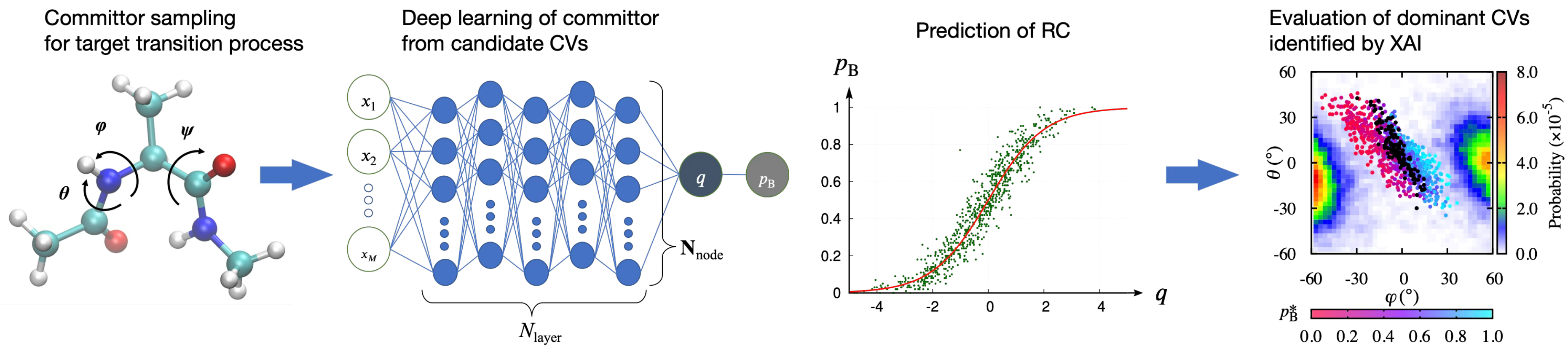}
\caption{
Schematic illustration of the explainable deep learning
framework for identifying RCs
based on committor $p_\mathrm{B}^*$ sampling.
First, 
MD trajectories are generated to sample the committor $p_\mathrm{B}^*$
 for the target transition process.
The candidate CVs, $x_i$, are utilized as input features and mapped 
 to the output $q$ by a neural network  model of $N_\mathrm{layer}$
 consisting of $\mathbf{N}_\mathrm{node}$.
The neural network is trained 
such that the committor $p_\mathrm{B}^*$ follows a sigmoidal function
 $p_\mathrm{B}(q) = [1 + \tanh(q)]/2$, where $q$ serves as the RC.
Finally, XAI analysis identifies the dominant CVs contributing to the RC
 $q$, enabling evaluation of the committor $p_\mathrm{B}^*$
 distribution on
 the corresponding free-energy landscape.
Reproduced from 
Mori \textit{et al.},
J. Chem. Phys. \textbf{153},
 054115 (2020) with the permission of 
AIP
 Publishing LLC.~\protect\cite{mori2020Learning} 
Reproduced with permission from 
Kawashima \textit{et al.}, APL Mach. Learn. \textbf{3}, 016113
 (2025).
Copyright 2025 Authors, licensed under a Creative
 Commons Attribution (CC BY) license.~\protect\cite{kawashima2025Investigating}
Reproduced with permission from 
Kikutsuji \textit{et al.}, J. Chem. Phys. \textbf{156}, 154108
 (2022).
Copyright 2022 Authors, licensed under a Creative Commons Attribution
 (CC BY) license.~\protect\cite{kikutsuji2022Explaining}
}
\label{fig:workflow}
\end{figure*}

Recently, we developed a deep learning framework trained on committor
values $p_\mathrm{B}^*$ of configurations sampled near the saddle point
region from MD simulations.~\cite{kikutsuji2022Explaining}
The schematic illustration of the work flow is shown in Fig.~\ref{fig:workflow}.
In this approach, candidate CVs are employed as input features to a
neural network, and the
corresponding RC is predicted as the output by training the 
network using $p_\mathrm{B}^*$ as the learning target.
The cross-entropy, which quantifies the discrepancy between the sigmoidal function,
$p_\mathrm{B}(q)=[1+\tanh(q)]/2$, and the pre-evaluated committor
values $p_\mathrm{B}^*$ that vary continuously from 0 to 1, is employed 
as a loss function and is minimized during
training.~\cite{mori2020Dissecting, mori2020Learning}
This formulation extends the likelihood maximization in 
that the cross-entropy is derived from the Kullback--Leibler (KL) divergence
by expressing the likelihood in logarithmic form.~\cite{mori2020Dissecting}

Because deep learning models often behave as black-boxes, eXplainable AI
(XAI) techniques are further employed to quantify the contribution of
each input variable to predictions.~\cite{adadi2018Peeking, molnar2020Interpretable}
XAI can be regarded as a class of model-agnostic interpretation methods for deep learning models.
As major implementations of XAI, we employed Local Interpretable
Model-agnostic Explanations (LIME)~\cite{ribeiro2016Whya} and the game theory–based framework
known as Shapley Additive exPlanations (SHAP)~\cite{lundberg2017unified}.
XAI approach enables the identification of CVs with significant
contribution, providing a representation of the TS on the free-energy
landscape that is consistent with the underlying transition pathways.
As an application, we investigated the isomerization reaction of alanine
dipeptide and identified specific dihedral angles as the primary
contributor to the RC.~\cite{kikutsuji2022Explaining}
We also demonstrated that the isomerization can also be characterized
by specific interatomic
distances.~\cite{okada2024Unveiling}
Moveover, we investigated how hyperparameters in the neural network
model influences the performance and reliability of the
identified RC.~\cite{kawashima2025Investigating}
Furthermore, 
we investigated the RC for NaCl ion dissociation–association process in water.
In particular, 
to describe the solvent environment surrounding the ions, we employed 
atom-centered symmetry functions (ACSFs), which are 
widely used as descriptors in machine learning
potentials,~\cite{behler2007Generalized, behler2011Atomcentered, behler2016Perspective} as
input CVs to the neural network.~\cite{okada2026Deep}

Other studies have provided interpretability for RCs
identified through committor based deep learning approaches. 
Frassek \textit{et al.} reported the application of an autoencoder
architecture comprising an encoder, a reconstruction decoder, and a
committor decoder to obtain low-dimensional representations of RCs from
a large set of input CVs.~\cite{frassek2021Extended}
Jung \textit{et al.} introduced a sampling method that determines
relevant RCs using maximum likelihood estimation combined with deep
learning, and further employed symbolic regression to derive
interpretable mathematical expressions representing trained models.~\cite{jung2023Machineguided} 
Neumann and Schwierz applied a related deep learning
framework to predict the RC governing magnesium binding to RNA and used
permutation importance to quantify feature relevance from the input CVs.~\cite{neumann2022Artificial}
Naleem \textit{et al.} investigated RCs for the committor
using 17 machine learning models, including Light Gradient Boosting
Machine (LGBM) based on the decision tree method~\cite{ke2017LightGBM}, and quantified input
CVs contributions with SHAP.~\cite{naleem2023Exploration}
Their results showed that the combination of LGBM and SHAP yielded the
most rubust description of alanine dipeptide isomerization.

This review introduces an explainable deep learning framework based on
the committor for
identifying the RC and its applications to complex molecular systems.
The structure of this review is as follows.
Section II outlines the background and methodology of our explainable
deep learning approach.
Section III presents the applications to isomerization of alanine
depeptide and ion-pair dissociation in water, and discusses
the hyperparameter tuning of the neural network.
Section IV provides a summary and an outlook.

\section{Methods}

\subsection{Committor as a measure for RC optimality}

Committor $p_\mathrm{B}^*(\mathbf{R})$ is defined as a probability to reach state B prior to reaching state A from a given coordinate $\mathbf{R}$ in a two-state system with states A and B.
A naive approach to calculate $p_\mathrm{B}^*$ is to perform multiple short MD simulations with velocities randomly generated from a Maxwell--Boltzmann distribution and count the probability of reaching state B.
$p_\mathrm{B}^*$ is 0 when $\mathbf{R}$ is near state A whereas it becomes 1 near state B.
At the TS, $p_\mathrm{B}^*$ is expected to be 0.5, \textit{i.e.}, there is equal chance of reaching states A and B.
In high dimensional systems, it is often convenient to describe the system with a RC $q$ that connects states A and B and construct a PMF along $q$.
Ideally, the committor as a function of $q$, $p_\mathrm{B}(q)$, is expected to smoothly change monotonically from 0 (at state A), pass 0.5 (at TS), and reach 1 (at state B).
Thus, committor can be used as a measure to evaluate the quality of the RC $q$.
The monotonic change of the committor as a function of a RC has
been modeled using several functions including the sigmoidal function
$\left[ 1 + \tanh(q) \right]/2$ and error function
$\mathrm{erfc}(-q)/2$.~\cite{peters2017Reaction, mori2020Dissecting}
While the sigmoidal function is more popular due to the relevance with
the neural network framework, the error function also has an advantage
for its relationship with the Kramers--Langer--Berezhkovskii--Szabo
theory.~\cite{peters2016Reaction, berezhkovskii2005Onedimensional,mori2020Learning}

In practice, $p_\mathrm{B}(\mathbf{R})$ along $q(\mathbf{R})$ do not follow the ideal curve unless the RC is chosen wisely.
Thus, from a set of points $\{k\}$ with $q^{(k)} = f(\mathbf{R}^{(k)})$ and
$p_\mathrm{B}^{*}(k) = p_\mathrm{B}^{*}(\mathbf{R}^{(k)})$,
the quality of the trial RC $q(\mathbf{R})$ can be examined using committor as a measure by comparing the deviation of the data points $p_\mathrm{B}^{*}(k)$ from the ideal curve $p_\mathbf{B}(q^{(k)})$.

\subsection{Cross-entropy minimization}

The probability distribution is defined as $P_\mathrm{B}(q) =
p_\mathrm{B}(q)/N$ where $N$ is the number of points ($N = \sum_{k}$).
The discrepancey between the pre-evaluated and ideal probabilities, denoted as $P_\mathrm{B}^*(q) (=p_\mathrm{B}^*(q)/N)$ and $P_\mathrm{B}(q) (=p_\mathrm{B}(q)/N)$, respectively, can be measured using the KL divergence $\mathcal{D}_\mathrm{KL}$,
\begin{align}
&  \mathcal{D}_{\mathrm{KL}}\left( P_\mathrm{B}^{*} || P_\mathrm{B} \right) \notag \\
=& \sum_{k} \sum_{\mathrm{I} = \mathrm{A}, \mathrm{B}} P_\mathrm{I}^{*}(k) \ln \dfrac{P_\mathrm{I}^{*}(k)}{P_\mathrm{I}\left( q^{(k)} \right)} \notag \\
=& \sum_k \left\{ P_\mathrm{B}^{*}(k) \ln \dfrac{P_\mathrm{B}^{*}(k)}{ P_\mathrm{B}\left( q^{(k)} \right)}
+ \left( 1/N - P_\mathrm{B}^{*}(k) \right) \ln \dfrac{ 1/N - P_\mathrm{B}^{*}(k)}{ 1/N - P_\mathrm{B} \left( q^{(k)} \right)} \right\} \notag \\
=& \dfrac{1}{N} \sum_k \left\{ p_\mathrm{B}^{*}(k) \ln \dfrac{p_\mathrm{B}^{*}(k)}{ p_\mathrm{B}\left( q^{(k)} \right)} \right.
\left. + \left( 1 - p_\mathrm{B}^{*}(k) \right) \ln \dfrac{ 1 - p_\mathrm{B}^{*}(k)}{ 1 - p_\mathrm{B} \left( q^{(k)} \right)} \right\} ,
\label{eq:KL_divergence}
\end{align}
where $q^{(k)}$ is a trial RC at point $k$, $p_\mathrm{B} \left( q^{(k)} \right)$ is the model function that describe the change of $p_\mathrm{B}$ along $q$, and $p_\mathrm{A}^{*}(k)$ and $p_\mathrm{A}\left(q^{(k)}\right)$ satisfy $p_\mathrm{A}^{*}(k) = 1 - p_\mathrm{B}^{*}(k)$ and $p_\mathrm{A}\left(q^{(k)}\right) = 1 - p_\mathrm{B}\left(q^{(k)}\right) $, respectively.
$D_{\mathrm{KL}}\left( p_\mathrm{B}^{*} || p_\mathrm{B} \right)$ is a measure to quantify the difference in the distribution of $p_\mathrm{B}^{*}$ from $p_\mathrm{B}$. 
$\mathcal{D}_{\mathrm{KL}}$ satisfies $\mathcal{D}_{\mathrm{KL}}\left( p_\mathrm{B}^{*} || p_\mathrm{B} \right) \ge 0$ and becomes 0 only when the two distributions match exactly.
Thus, $\mathcal{D}_{\mathrm{KL}}=0$ for an ideal RC, and finding an ideal RC is equivalent to minimizing Eq.~\eqref{eq:KL_divergence}.

$\mathcal{D}_{\mathrm{KL}}$ can be divided into the self- and
cross-entropy terms ($\mathcal{H}_\mathrm{S}$ and
$\mathcal{H}_\mathrm{X}$, resepectively)
\begin{align}
  \mathcal{D}_{\mathrm{KL}}\left( p_\mathrm{B}^{*} || p_\mathrm{B} \right)
  =&  - \dfrac{1}{N} \mathcal{H}_S(p_\mathrm{B}^{*}) + \dfrac{1}{N} \mathcal{H}_\mathrm{X}(p_\mathrm{B}^{*}, p_\mathrm{B}), 
\end{align}
where
\begin{align}
  \mathcal{H}_\mathrm{S}\left( p_\mathrm{B}^{*} \right) =&
   - \sum_k p_\mathrm{B}^{*}(k) \ln p_\mathrm{B}^{*}(k) \notag \\
  &- \sum_k \left[ 1 - p_\mathrm{B}^{*}(k) \right] \ln\left[ 1 - p_\mathrm{B}^{*}(k) \right] 
\end{align}
and 
\begin{align}
   \mathcal{H}_\mathrm{X}\left( p_\mathrm{B}^{*}, p_\mathrm{B} \right) =&
   - \sum_k p_\mathrm{B}^{*}(k) \ln p_\mathrm{B} \left( q^{(k)} \right) \notag \\
  &- \sum_k \left[ 1 - p_\mathrm{B}^{*}(k) \right] \ln\left[ 1 - p_\mathrm{B} \left( q^{(k)} \right) \right].
\label{eq:def_entropy}
\end{align}
As the self-entropy term is invariant to changes in the trial RC $q$,
minimizing $\mathcal{D}_{\mathrm{KL}}$ is equivalent to
minimizing the cross-entropy term.
The following objective function $\mathcal{L}$ is thus defined to
quantify the difference between the modeled and pre-evaluated committors:
\begin{align}
\mathcal{L}(q) = \mathcal{H}_\mathrm{X}(p_\mathrm{B}^{*}, p_\mathrm{B}(q)) + \mathbf{\lambda} G(q). \label{eq:L_q}
\end{align}
Note that the first term is a generalization of the log-likelihood
maximization.~\cite{peters2006Obtaining, peters2007Extensions}
The second term describes the penalty function applied to $q$ so as to
suppress overfitting and $\mathbf{\lambda}$ is the regularization
parameter that controls the penalty term. 
An example is the $L_2$ norm regularization.~\cite{mori2020Learning}
By minimizing Eq.~\eqref{eq:L_q}, a RC $q$ that most closely mimic the
monotonic behivior, \textit{e.g.}, $p_\mathrm{B}(q) = \left[ 1 +
\tanh(q)\right]/2$, can be obtained.

\subsection{Neural network model for predicting RC}

The RC $q$ is obtained as a function of the coordinate $\mathbf{R}$.
However, the mapping form $\mathbf{R}$ to $q$ is nontrivial.
In many cases, the system $\mathbf{R}$ is first described with pre-selected set of CVs.
The RC is defined as a function of the CVs, either in a linear or nonlinear form.
Deep neural network (DNN) model is a nonlinear mapping approach to obtain the RC from the candidate CVs.
In DNN, the candidate CVs, discussed in Section~\ref{sec:CV}, serves as
the input features ($\mathbf{x}=(x_1, x_2, \cdots, x_M)$ with $M$ being
the number of input variables) after standardization, and $q$ is obtained as the output.
A common model is the multilayer perceptron model, which consists of $N_{\mathrm{layers}}$ hidden layers and $\mathbf{N}_{\mathrm{node}}$ nodes (Fig.~\ref{fig:workflow}).
Further choices of the model includes the activation function (the leaky rectified linear unit (Leaky ReLU) or scaled ExponentialLinear Unit), and regularizations ($L_1$, $L_2$, etc).
These parameters, \textit{i.e.} hyperparameters, needs to be chosen prior to optimizing the DNN model, and choosing the hyperparameters are unfortunately nontrivival and highly tedious.
Here we only consider leaky ReLU with a leaky parameter set of 0.01, and utilize $L_2$ regularization.
The number of hidden layers were also fixed to five with 400, 200, 400, 200, and 400 nodes,
and the dropout at a rate of 0.5 was set to every layer in Secs.~\ref{sec:ala2} and \ref{sec:ion}.
The effect of hyperparameters to the model result are discussed in Section~\ref{sec:hyperparameter}.

\subsection{Input features as candidate CVs}
\label{sec:CV}

The construction of CVs used as input features is a
crucial aspect of committor based machine learning for identifying RCs.
In the study by Ma and Dinner on the isomerization of alanine dipeptide,
pairwise distances, bond angles, and dihedral angles, including improper
dihedrals, were employed as inputs to a genetic neural network.~\cite{ma2005Automatic}
Furthermore, to investigate the RC for isomerization in explicit
solvents, it is necessary to incorporate CVs that explicitly describe
solute-solvent and solvent-solvent interactions. 
In particular, CVs such as the solvent accessible surface area, radius
of gyration, and excluded volume of the solute, as well as density
fields defined on spherical polar coordinate grids and energetic, force,
and torque components within the solute molecule, have been considered.~\cite{ma2005Automatic}
Section~\ref{sec:ala2} outlines our deep learning study of the
isomerization reaction of alanine dipeptide. 
We considered 45 dihedral angles, including improper ones. 
To account for the periodicity of dihedral angles, both cosine and sine
forms were used, generating a total of 90 CVs.
Section~\ref{sec:hyperparameter} further describes our study of neural network
hyperparameter tuning using the alanine dipeptide in water system. 
In that work, the electrostatic and van der Waals potentials from solvents to solute atoms were
included in addition to dihedral angles, resulting in a total of 134 CVs.

Another complex molecular system in which solvent effects play a central
role is gas hydrate nucleation, for which 
a variety of CVs have been examined. 
The nucleus size is commonly quantified using the mutually coordinated
guest parameter, which counts the number of gas molecules
participating in the largest solid nucleus in the system.~\cite{barnes2014Twocomponent} 
In addition, the number of water molecules incorporated into the
nucleus and the cage types of the growing nucleus are also employed as CVs.~\cite{arjun2019Unbiased}
Other CVs include the bond-orientational order parameters proposed by
Steinhardt~\cite{steinhardt1983Bondorientationala}, and their
variants~\cite{lechner2008Accurate} were used in similar liquid-solid
nucleation systems.~\cite{beckham2011Optimizing}

Furthermore, the CV construction becomes challenging in solvent-mediated
ion pair dissociation.
M\"{u}llen \textit{et al.} applied the inertial likelihood maximization
method,~\cite{peters2012Inertial} an extension of the likelihood maximization
that incorporating the velocity, acceleration, and jerk of the inter-ion
distance $r_\mathrm{ion}$,
to NaCl ion pair dissociation in water.~\cite{mullen2014Transmission}
In this approach, 
a total of 71 candidate CVs representing
water bridging structures, namely, configurations in which the O and H
atoms of a water molecule simultaneously coordinate to the Na and Cl
ions, respectively, were considered.
It was demonstrated that the inter-ionic water density $\rho$ and 
and the number of bridging waters $N_\mathrm{B}$
contribute most significantly to the RC.
Specifically, the inter-ionic water density $\rho$ is defined as
\begin{equation}
\rho = \left(\frac{1}{2\pi\sigma^2}\right)^{3/2}
\sum_{w}
\exp\left(
-\frac{| \mathbf{r}_w - \mathbf{r}_\mathrm{mid}|^2}{2\sigma^2}
\right),
\end{equation}
where $\mathbf{r}_w$ denotes the center-of-mass position of the $w$-th water
molecule and $\mathbf{r}_\mathrm{mid}$ is the midpoint between the Na and Cl
ions. 
The parameter $\sigma$ controls the effective volume over which water
molecules are counted and is chosen as $r_\mathrm{ion}/2$.
The number of bridging waters $N_\mathrm{B}$ is evaluated in two steps. 
First, the ion–water coordination function is defined as
\begin{equation}
f_{s-w} = \frac{1 - \tanh\left[a (R_\mathbf{s-w} - b)\right]}{2},
\end{equation}
where the subscript $s$ denotes the ion species (Na or Cl). 
The distance $R_\mathbf{s-w}$ corresponds to the distance between the Na ion
and the $w$-th water O atom for Na coordination, and between the Cl ion
and the $w$-th water H atom for Cl coordination. 
The parameters are set to $a = 3$ {\AA}$^{-1}$ and $b = 3.2$ {\AA}.
The number of water molecules simultaneously coordinating both ions is
then given by
\begin{equation}
N_\mathrm{B} = \sum_{w} \min \left(f_{\mathrm{Na}-w}, f_{\mathrm{Cl}-w}\right).
\end{equation}
In particular, $\rho$ and $N_\mathrm{B}$ were identified as
important CVs, together with the interion distance $r_\mathrm{ion}$, through
analyses of the PMF.~\cite{mullen2014Transmission,
yonetani2015Distinct,yonetani2017Solventcoordinate, salanne2017Ca2+Cl,
joswiak2018Ion, oh2019Understanding, zhang2020Dissociation,
wang2022Influence, wilke2025NaCl} 
These studies showed that the formation of water bridges lowers the
dissociation free energy barrier and facilitates ion pair dissociation.

Jung \textit{et al.} recently employed ACSFs as CVs to represent the
solvent environment 
in studies of ion pair dissociation.~\cite{jung2023Machineguided}
ACSFs are commonly used as descriptors in machine learning
potentials,~\cite{behler2007Generalized, behler2011Atomcentered,
behler2016Perspective} and 
quantify the distribution of neighboring atoms at specific distances and
angles around a reference atom.~\cite{behler2011Atomcentered} 
Because they are invariant under translational and
rotational operations, ACSFs provide a systematic 
representation of local solvent environment.
To represent the solvent environment around the ions, 
two types of ACSFs, $G_i^2$ and $G_i^5$ for
reference atom $i$, were employed as input variables for neural networks.~\cite{jung2023Machineguided}
Here, these ACSFs are defined as
\begin{align}
G_i^{2, Z_1} &= \sum_{j\ne i}^{|Z_1|} e^{-\eta(R_{ij}-R_\mathrm{s})^2} \cdot
 f_\mathrm{c}(R_{ij}), \\
G_i^{5, Z_1, Z_2} &= 2^{1-\zeta} \sum_{j\ne i}^{|Z_1|} \sum_{k\ne
 i}^{|Z_2|} (1+\lambda \cos \theta_{ijk})^\zeta \nonumber\\
&\qquad \cdot e^{-\eta((R_{ij}-R_\mathrm{s})^2+(R_{ik}-R_\mathrm{s})^2)} 
 \cdot f_\mathrm{c}(R_{ij}) \cdot
 f_\mathrm{c}(R_{ik}), 
\end{align}
respectively.
Note that $R_{ij}$ ($R_{ik}$) represents the distance between atoms $i$
and $j$ ($k$), and 
$\theta_{ijk}$ denotes the angle formed by atoms $j$ and $k$ with 
atom $i$ at the center.
$Z_1$ and $Z_2$ represent atomic species
surrounding the reference atom 
and the absolute value notation denotes their
respective atomic numbers.
Furthermore, the cut-off function is introduced as
\begin{equation}
f_\mathrm{c} (R_{ij}) = 
\begin{cases}
0.5\cdot \left[\cos(\pi R_{ij}/R_\mathrm{c})+1\right]  
& \text{for $R_{ij} < R_\mathrm{c}$} \\
0 
& \text{for $R_{ij} > R_\mathrm{c}$},
\end{cases}
\end{equation}
with the cut-off radius, $R_\mathrm{c}$.

The ACSF $G_i^2$
is defined as the sum of a Gaussian function weighted by the cutoff
function. 
The width of the Gaussian is determined by the parameter $\eta$, 
while its center can be positioned at a specific radial distance using
the parameter $R_\mathrm{s}$.
This function is particularly suitable for characterizing the spherical
shell environment surrounding a reference atom.
The ACSF $G_i^5$ extends $G_i^2$ by incorporating an additional target
atom and including an angular component. 
The parameter $\lambda$ can take values of $1$ or $-1$, shifting the
maximum of the cosine function to $\theta_{ijk}=0^\circ$ and
$\theta_{ijk}=180^\circ$, respectively.
The parameter $\zeta$
controls the angular resolution, with larger values of $\zeta$
corresponding to a narrower range of nonzero contributions in the ACSF.
Section~\ref{sec:ion} outlines our explainable deep learning framework
for ion pair dissociation, in which a total of 1,296 ACSFs are used as CVs.
The correlations between
water bridging structures described
by $\rho$ and $N_\mathrm{B}$ and the ACSFs identified by the SHAP
analysis are further discussed.

\begin{figure*}[t]
\centering
\includegraphics[width=0.9\textwidth]{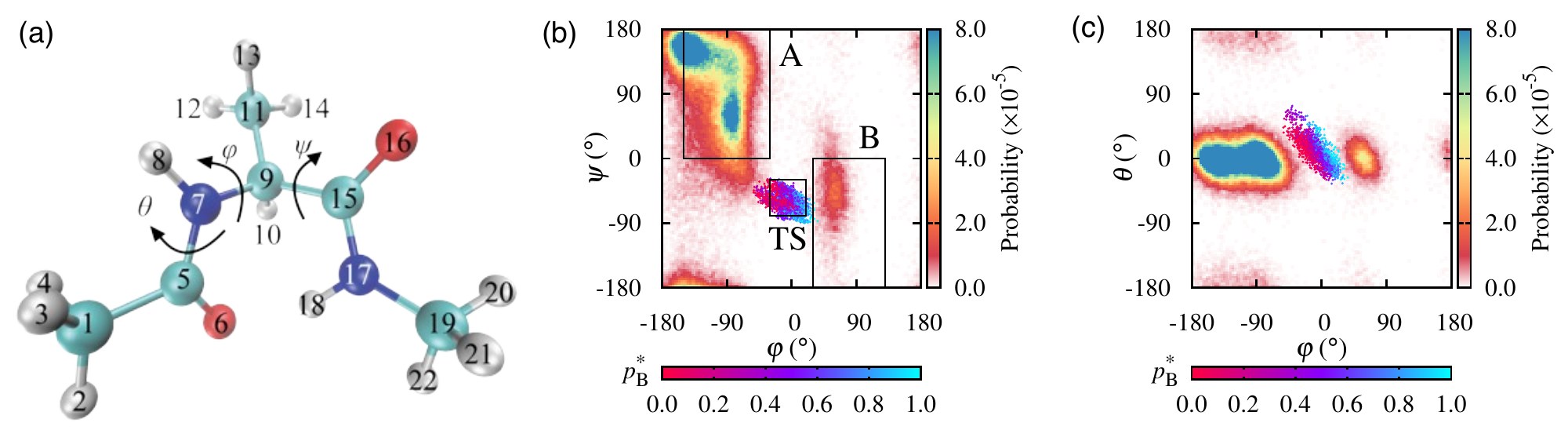}
\caption{
(a) Atomic indices assigned to alanine dipeptide. 
The three major dihedral angles, $\varphi$, $\psi$, and $\theta$,
 are indicated.
(b) Probability distribution in the $(\varphi, \psi)$ 
plane. 
The dashed lines define state A 
$[(-150^\circ, 0^\circ )\le (\varphi, \psi) \le (30^\circ, 180^\circ)]$, 
state B 
$[(30^\circ, -180^\circ )\le
 (\varphi, \psi) \le (130^\circ, 0^\circ )]$, 
and the intermediate region TS 
$[(-30^\circ, -80^\circ ) \le (\varphi, \psi)\le (20^\circ,-30^\circ)]$.
(c) Probability distribution in the $(\varphi, \theta)$ plane.
In (b) and (c), points are colored according to the committor value $p_\mathrm{B}^*$.
as indicated by the color bar.
Reproduced with permission from 
Kikutsuji \textit{et al.}, J. Chem. Phys. \textbf{156}, 154108
 (2022).
Copyright 2022 Authors, licensed under a Creative Commons Attribution
 (CC BY) license.~\protect\cite{kikutsuji2022Explaining}
}
\label{fig:Ala2}
\end{figure*}

\begin{figure}[t]
\centering
\includegraphics[width=0.45\textwidth]{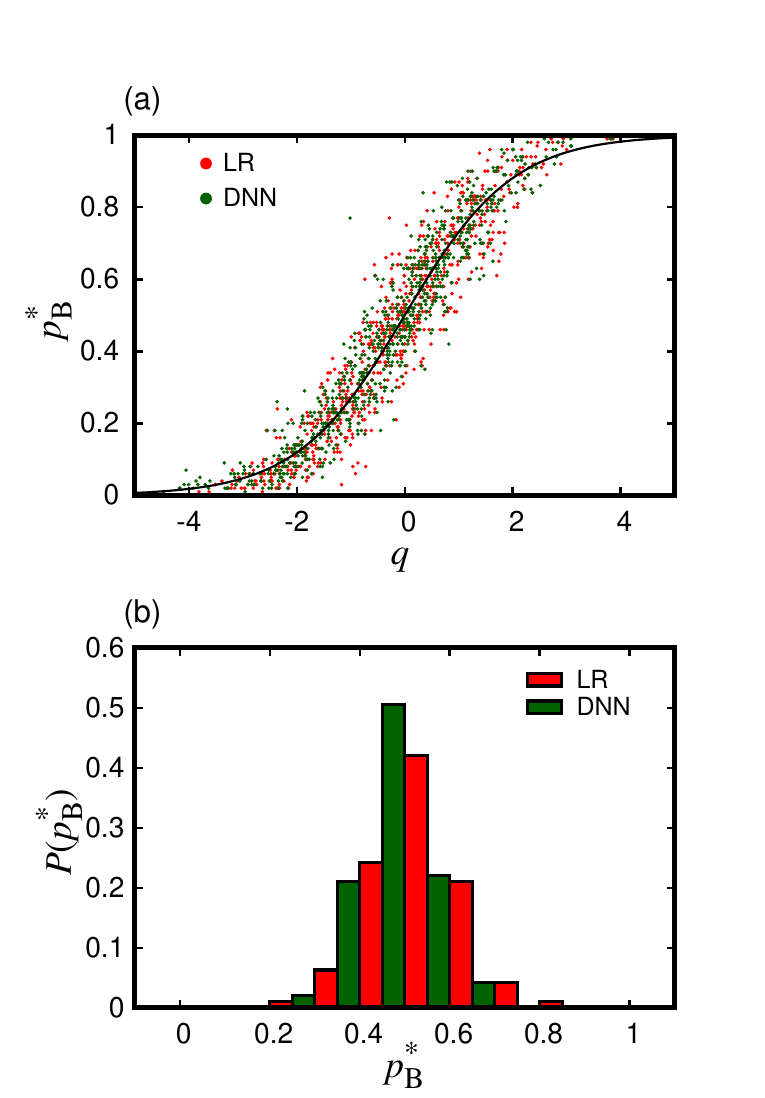}
\caption{(a) Relationship between the committor value
 $p_\mathrm{B}^*$ and the RC $q$ predicted by LR (red) and DNN (green)
 for the test dataset (800 points). 
The black solid line denotes the sigmoidal function  
$p_\mathrm{B}(q) = [1+ \tanh(q)]/2$. 
(b) Probability distribution of the committor $p_\mathrm{B}^*$ obtained
 from LR (red) and DNN (green) within the range $-0.2 < q < 0.2$ shown in (a).
Reproduced with permission from 
Kikutsuji \textit{et al.}, J. Chem. Phys. \textbf{156}, 154108
 (2022).
Copyright 2022 Authors, licensed under a Creative Commons Attribution
 (CC BY) license.~\protect\cite{kikutsuji2022Explaining}
}
\label{fig:committor}
\end{figure}

\begin{figure*}[t]
\centering
\includegraphics[width=0.9\textwidth]{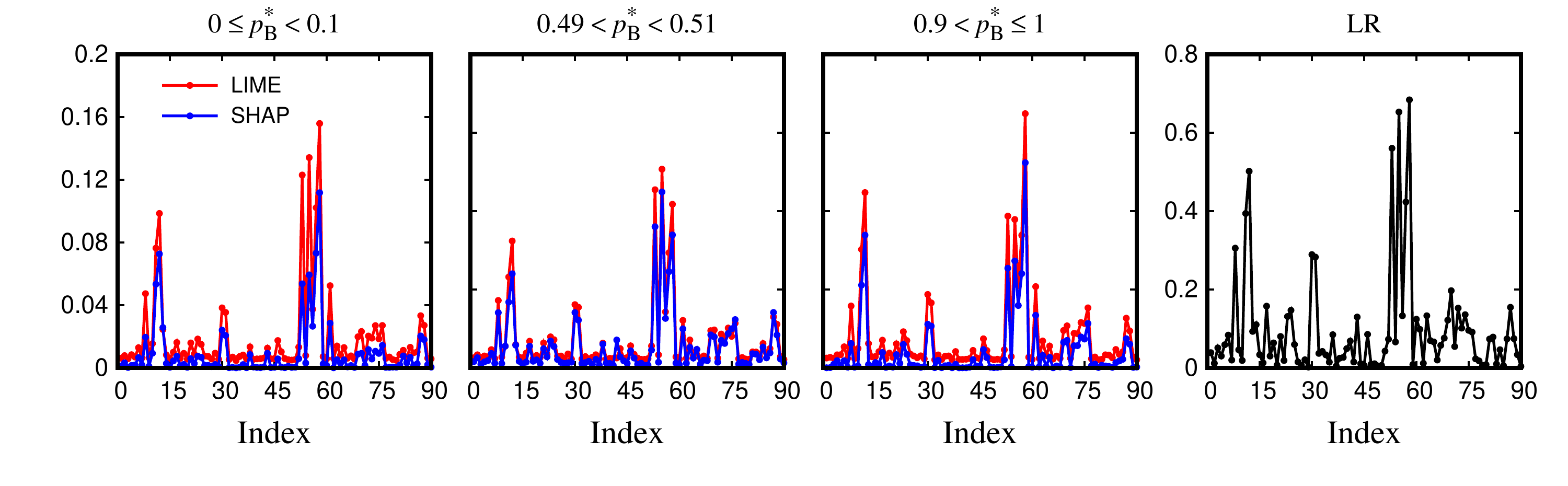}
\caption{Feature contribution of each CV in absolute value obtained
using LIME (red) and SHAP (blue). 
From left to right: $0 \le p_\mathrm{B}^* < 0.1$ (near state A), $0.49 <
 p_\mathrm{B}^* < 0.51$ (near TS), and $0.9 < p_\mathrm{B}^* \le 1$
 (near state B). 
The rightmost panel shows the absolute values of the coefficients $(w_1,
 w_2, \cdots, w_{90})$ obtained from the LR model.
Reproduced with permission from 
Kikutsuji \textit{et al.}, J. Chem. Phys. \textbf{156}, 154108
 (2022).
Copyright 2022 Authors, licensed under a Creative Commons Attribution
 (CC BY) license.~\protect\cite{kikutsuji2022Explaining}
}
\label{fig:XAI}
\end{figure*}

\subsection{XAI analysis}

XAI was applied to interpret committor predictions obtained from neural
network training. 
Because deep-learning models are highly complex, direct identification
of interpretable features is difficult. 
To address this issue, we employed XAI techniques, LIME and SHAP, to
quantify the contribution of each input variable to the predictions of
the black-box classifier through model-agnostic interpretable models.

LIME enables interpretation of the contribution of individual input
variables in a black-box neural network through an interpretable
surrogate model.~\cite{ribeiro2016Whya}
Specifically, LIME constructs a local explanation model based on
linear regression that approximates the behavior of the original model
in the vicinity of the data point of interest. 
In this context, input
variables associated with large regression coefficients provide insight
into the prediction.
The interpretable feature $\xi(\mathbf{x})$
is obtained by fitting a linear regression model $g$
to the predictions of the black-box model $f$, represented by the neural
network, for the input data $\mathbf{x}$.
This procedure is formulated as
\begin{equation}
\xi(\mathbf{x}) = \text{argmin}_{g\in G}\left(L(f, g, \pi_x)+\Omega(g)\right),
\label{eq:LIME}
\end{equation}
where 
$L(f,g,\pi_x)$ 
denotes a loss function that quantifies the discrepancy between the predictions of $f$
and $g$, and $G$ represents the class of candidate explanation models. 
The regularization term $\Omega(g)$ penalizes model complexity to promote interpretability. 
Specifically, $L(f,g,\pi^\text{LIME}_x)$ is defined as
\begin{equation}
L(f, g, \pi^\text{LIME}_x)  = \sum_{\mathbf{z}, \mathbf{z'} \in Z}
 \pi^\text{LIME}_x(\mathbf{z}) (f(\mathbf{z})-g(\mathbf{z'}))^2,
\end{equation}
where $\pi^\text{LIME}_x$ denotes the proximity measure characterizing 
the locality around the input data $\mathbf{x}$ to be explained.

In practice, 
perturbed samples $\mathbf{z}$
generated around 
$\mathbf{x}$ and weighted by the proximity measure are mapped onto binary
variables $\mathbf{z'}\in \{0, 1\}^M$, where
$M$ denotes the number of input variables, to provide a
human-interpretable representation, because the components of the
original input space are not always directly interpretable.
The proximity measure is further given by
\begin{equation}
\pi^\text{LIME}_x (\mathbf{z})=\exp(-D(\mathbf{x},\mathbf{z})^2/\sigma^2), 
\label{eq:LIME_weight}
\end{equation}
where 
$D$ is a distance metric and $\sigma$
controls the width of the neighborhood. 
By minimizing the object function, LIME yields a local surrogate 
model $g$
that approximates the behavior of $f$
near $\mathbf{x}$, from which the interpretable feature $\xi(\mathbf{x})$
is obtained.

Furthermore, SHAP, a game-theory–based method, was applied. 
Whereas LIME assumes that local model behavior can be approximated by
linear regression, SHAP enforces an additive decomposition in which the
sum of individual feature contributions equals the predicted value.~\cite{lundberg2017unified}
In SHAP, an additive feature attribution method provides a linear
function consisting of binary variables:
\begin{align}
g(\mathbf{z}') = \phi_0 + \sum_{i = 1}^M \phi_i z'_i.
\label{eq:additive_feature}
\end{align}
Here, $\phi_0$ denotes the bias term, and $\phi_i$
represents the contribution of the $i$-th feature $x'_i$, which is an
interpretable representation corresponding to the original input
variable $x_i$.
SHAP determines the coefficients $\phi_i$ using Shapley values, which
originate from game theory and allocate contributions among features.~\cite{shapley195317}
Note that LIME can be understood as an additive feature attribution method that provides
linear models in the binary vector space in the sense that 
the explanation model $g$ of LIME can be expressed by 
Eq.~(\ref{eq:additive_feature}).
Exact computation of Shapley values requires evaluation over all
permutations of feature inclusion, which is generally computationally
demanding. 
To address this limitation, Kernel SHAP was introduced to approximate
the Shapley values efficiently. 
In Kernel SHAP, the coefficients $\phi_i$ are obtained by solving a
weighted linear regression problem, thereby quantifying the contribution
of each feature to the model prediction.

Specifically, for Kernel SHAP, the following equations are used 
with Eq.~(\ref{eq:LIME}) of the LIME algorithm:
\begin{align}
\Omega(g) &=0,\label{eq:SHAP_penalty}\\
\pi_{{x}}^\text{SHAP}(\bm{z}')&=\frac{M-1}{{}_M
 C_{|\mathbf{z}'|}|\mathbf{z}'|(M-|\bm{z}'|)},\label{eq:SHAP_weight}\\
L(f, g, \pi_{{x}}^\mathrm{SHAP}) & =\sum_{\mathbf{z}'\in
 Z}\pi^\text{SHAP}_{x}(\bm{z}')\left(f(h_{{x}}^{-1}(\mathbf{z}'))-g(\mathbf{z}')\right)^2,
\end{align}
where $|\mathbf{z}'|$ is the number of non-zero features in $\mathbf{z}'$ and
$h_{{x}}$ is a mapping function of binary variables $\mathbf{z}'$
into the original input data $\mathbf{x}$.
The local explanation model $g(\mathbf{z}')$ approximating
$f(h_{{x}}^{-1}(\mathbf{z}'))$ can be obtained using a weighted
linear regression with Eq.~(\ref{eq:SHAP_weight}), which differs 
from the weight function used in LIME, as shown in Eq.~\eqref{eq:LIME_weight}.
Note the contrast between Eqs.~\eqref{eq:LIME_weight} and
\eqref{eq:SHAP_weight}, which are used in LIME and SHAP, respectively.
In practice,
if we use SHAP's kernel of Eq.~\eqref{eq:SHAP_weight} as LIME's kernel,
LIME will provide values similar to SHAP values.

The choice between LIME and SHAP, or the consistent use of both, merits
consideration. 
SHAP, developed as an additive feature attribution method and closely
related to LIME, and 
the features identified as having large
contributions by LIME and by SHAP are therefore expected to be consistent.
Because LIME generates perturbed samples randomly, the resulting
explanations may vary between runs.
This variability can be reduced by 
increasing the number of
generated perturbed samples.
SHAP offers a theoretical foundation grounded in game
theory.
By decomposing the predicted value into contributions that are
fairly allocated among features, SHAP provides a more sophisticated 
quantification of individual contributions than LIME.
However, the
computational cost of SHAP increases exponentially with the number of
features since the evaluation of Shapley values requires consideration of
many feature combinations.

%
%
\begin{figure}
\includegraphics[width=0.35\textwidth]{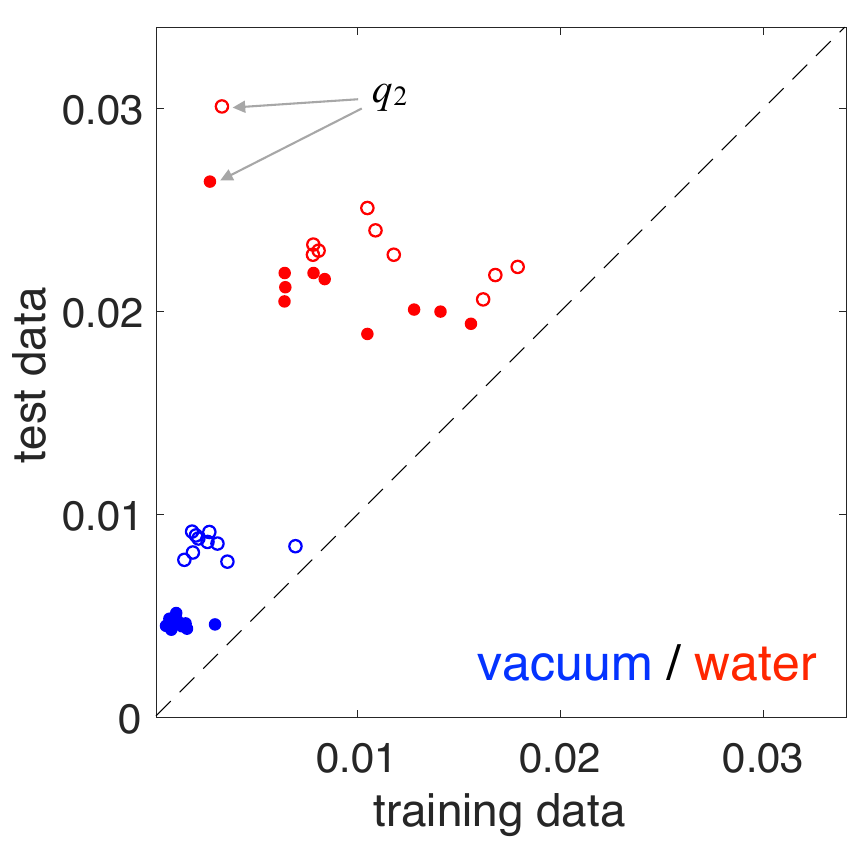}
\caption{Scatter plot for RMSEs between the predicted and reference
 {$p_\mathrm{B}$} for the training and test data sets. Filled and open
 circles indicate the RMSEs using full points and those at about the TS
 ($-0.2 < q_i < 0.2$), respectively. Blue and red colors are the results
 in vacuum and water.  $q_2$ in water shows a sign of slight
 overfitting, yet the RMSE against the test data is only marginally
 larger than the other RCs. 
Reproduced with permission from 
Kawashima \textit{et al.}, APL Mach. Learn. \textbf{3}, 016113
 (2025).
Copyright 2025 Authors, licensed under a Creative
 Commons Attribution (CC BY) license.~\protect\cite{kawashima2025Investigating}
}
\label{fig:rmse_train_test}
\end{figure}

%
%
\begin{figure}
\includegraphics[width=0.45\textwidth]{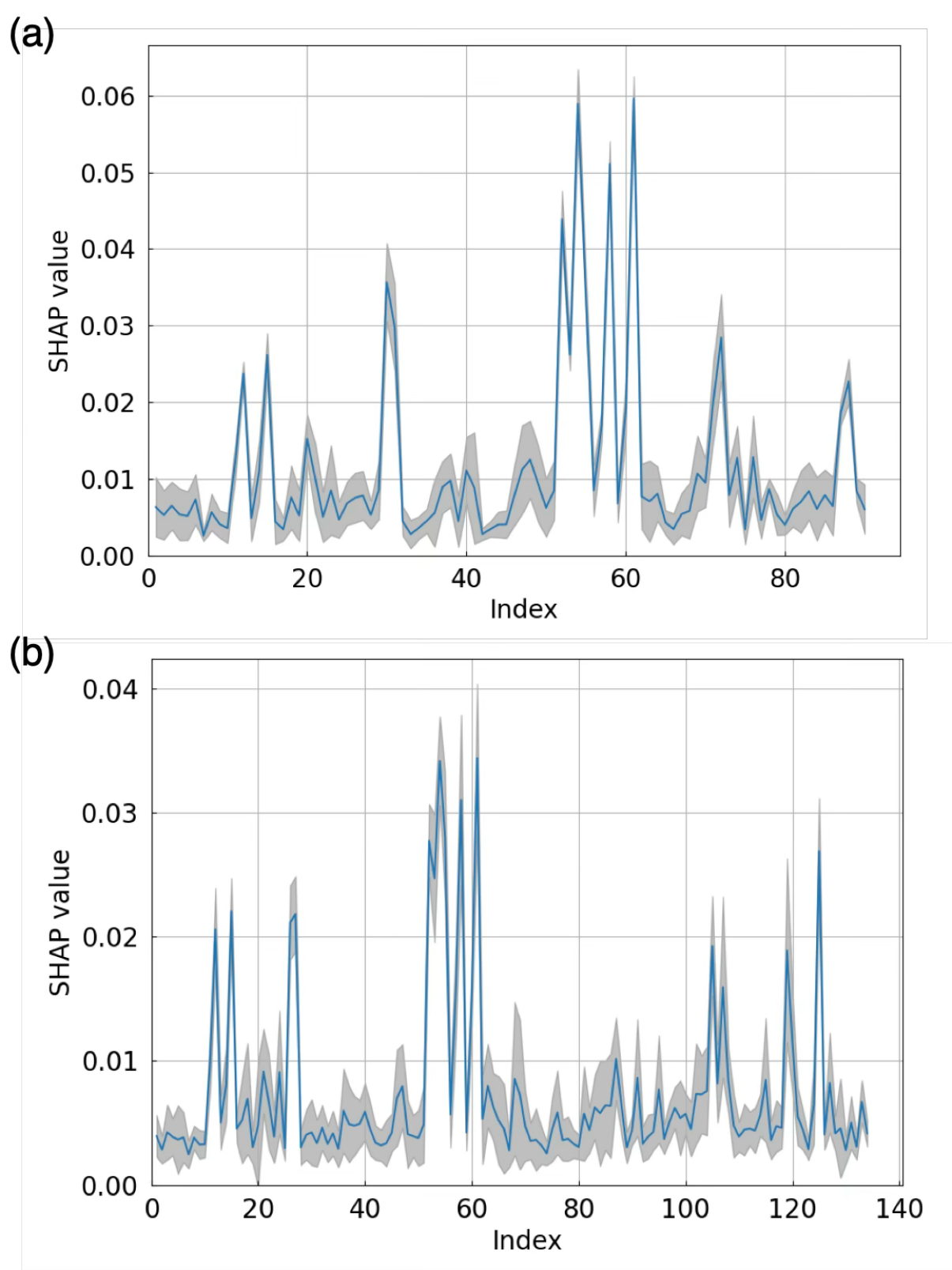}
\caption{Feature contributions of each CV in absolute value using SHAP
 for cases in (a) vacuum and (b) water. Blue lines and gray shades
 denote the average and variance calculated from the ten optimized RCs. 
Reproduced with permission from 
Kawashima \textit{et al.}, APL Mach. Learn. \textbf{3}, 016113
 (2025).
Copyright 2025 Authors, licensed under a Creative
 Commons Attribution (CC BY) license.~\protect\cite{kawashima2025Investigating}
}
\label{fig:shap_vac_wat}
\end{figure}

%
%
\begin{figure}
\includegraphics[width=0.45\textwidth]{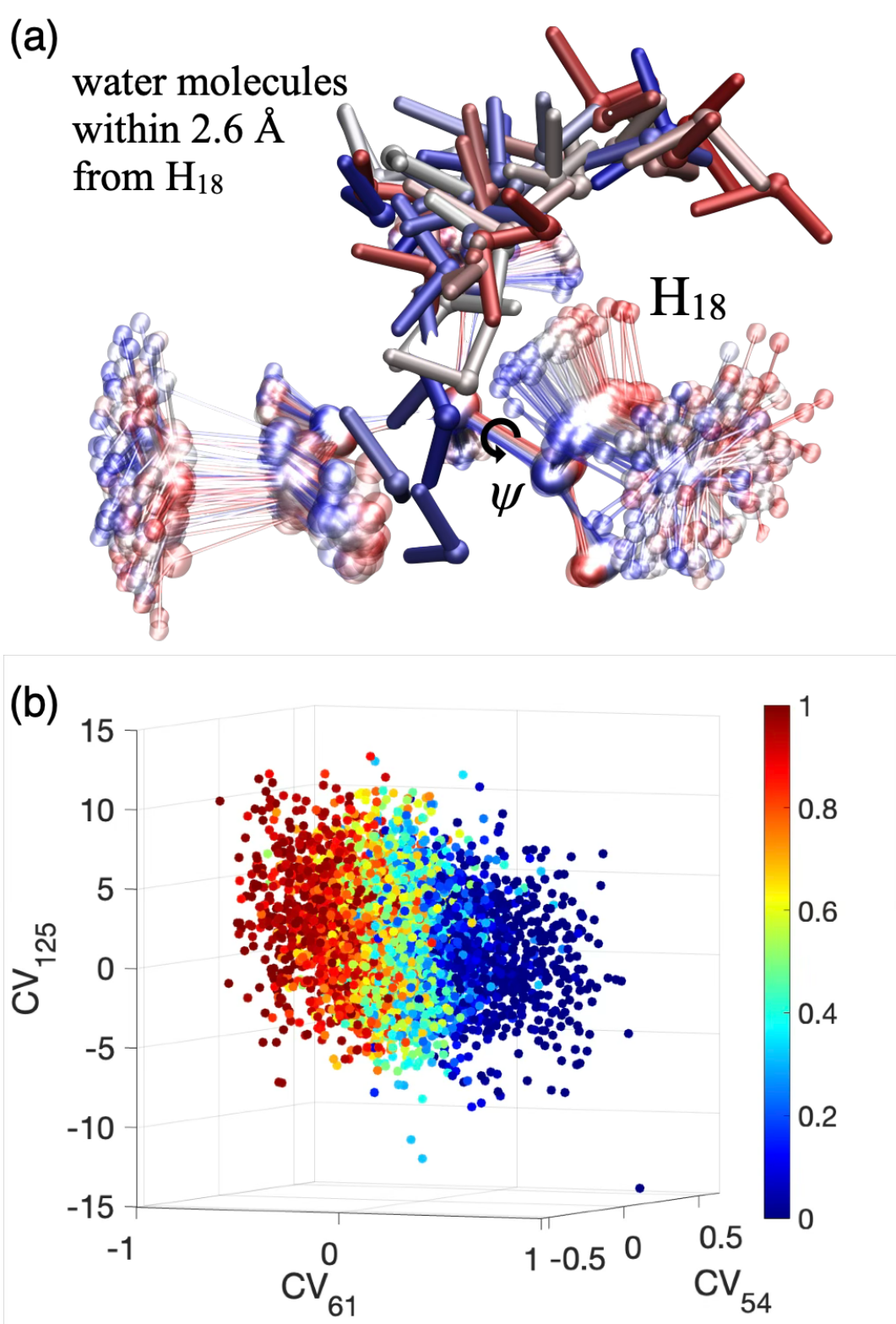}
\caption{(a) Structures within $\left|q_1\right| < 0.2$ sorted based on $\psi$ and aligned against the backbone atoms of alanine. RWB color range indicate the change of $\psi$ in the structures from -123$^\circ$ to -50$^\circ$. Water molecules within 2.6 {\AA} from H${}_{18}$ is shown.
Alanine dipeptide and water molecules are shown in blurry and sharp representations, respectively.
(b) Correlations between the changes in CVs and $p_\mathrm{B}$ in water. 
Reproduced with permission from 
Kawashima \textit{et al.}, APL Mach. Learn. \textbf{3}, 016113
 (2025).
Copyright 2025 Authors, licensed under a Creative
 Commons Attribution (CC BY) license.~\protect\cite{kawashima2025Investigating}
}
\label{fig:correlation_3d}
\end{figure}

\section{Applications}

\subsection{Alanine depeptide}
\label{sec:ala2}

Bolhuis \textit{et al.} applied the committor analysis to the
isomerization reaction of alanine
dipeptide~\cite{bolhuis2000Reaction}. 
The Ramachandran plot, which represents the probability distribution of
the backbone dihedral angles $\varphi$ and $\psi$, is commonly used to
characterize protein conformation (see Figure~\ref{fig:Ala2}(a) for
the definitions of $\varphi$ and $\psi$).~\cite{ramachandran1968Conformation}
This plot can be regarded as the PMF defined in terms of two CVs,
$\varphi$ and $\psi$.
In vacuum, two energetically stable conformations are observed: the 
$\beta$-sheet structure (state A) and the left-handed 
$\alpha$-helix structure (state B) (see Figure~\ref{fig:Ala2}(b) for the
definitions of states A and B). 
However, analysis of the committor distribution obtained by sampling initial configurations 
$x$ in the vicinity of the saddle point
showed that the distribution does not exhibit a peak at $p_\mathrm{B}^*=1/2$,
but instead is nearly uniform over the interval from 0 to 1. 
This behavior indicates that the saddle point on the PMF does not
necessarily correspond to the TS. 
They further demonstrated that inclusion of an additional dihedral angle $\theta$
is required for the committor distribution to develop a peak at $p_\mathrm{B}^*=1/2$
(see Figure~\ref{fig:Ala2}(a) for the definition of 
$\theta$). 
Thus, although the Ramachandran plot based on $\varphi$ and $\psi$
distinguishes states A and B, 
incorporation of $\theta$ is necessary to represent the RC appropriately.

We applied the explainable deep learning framework to 
the isomerization reaction of alanine dipeptide in vacuum.
State A was defined by $(-150^\circ,0^\circ)\le
(\varphi, \psi)\le
(30^\circ,180^\circ)$, whereas state B was defined by 
$(30^\circ,-180^\circ )\le (\varphi,\psi)\le (130^\circ,0^\circ)$.
An intermediate region corresponding to the expected TS, 
$(-30^\circ,-80^\circ )\le (\varphi, \psi)\le (20^\circ,-30^\circ )$,
was also defined, from which 2,000 initial configurations were sampled
using the aimless shooting. 
For each configuration, velocities were assigned randomly according to
the Maxwell--Boltzmann distribution at 300 K, and 100 trajectories of 1
ps duration were generated to evaluate the committor value
$p_\mathrm{B}^*$.
For each initial configuration, a total of 90 CVs
($M=90$) were constructed by including both cosine and sine
representations of all 45 dihedral angles in the molecule (see
Table I of Ref.~\onlinecite{kikutsuji2022Explaining} for the correspondence between CV indices and
dihedral angles). 
Figures~\ref{fig:Ala2}(b) and \ref{fig:Ala2}(c) show the distributions
of the sampled configurations, colored by the committor value
$p_\mathrm{B}^*$.
Figure~\ref{fig:Ala2}(c) displays the distribution obtained when
$\varphi$ and $\theta$
were chosen as the CVs. 
While configurations with $p_\mathrm{B}^*\approx 1/2$
are broadly dispersed on the $(\varphi, \psi)$ plane, 
a more distinct separatrix line emerges on the $(\varphi, \theta)$
plane. 
This observation indicates that $\theta$, rather than $\psi$,
contributes more directly to the RC, in agreement with
previous findings reported by Bolhuis \textit{et al}.~\cite{bolhuis2000Reaction}

A dataset comprising dihedral angles and committor values from 2,000
initial configurations was divided into training, validation, and test
sets in a 5:1:4 ratio. 
For comparison, linear regression (LR) implemented through a single-layer
perceptron was also carried out, where $q$ is expressed as
$q=w_0+\sum_{m=1}^M w_m x_m$ with coefficients $(w_0, w_1, \cdots,
w_M)$, 
including the bias term $w_0$.
All other calculation conditions are described in the original
paper.~\cite{kikutsuji2022Explaining}


The learned models of LR and DNN predict the relationship
between the committor value $p^{*}_\mathrm{B}$ and the RC $q$ from the
test dataset (800 samples). 
Figure~\ref{fig:committor}(a) shows the committor value
$p^{*}_\mathrm{B}$ as a function of $q$ for the test dataset. 
In both cases, $p^{*}_\mathrm{B}$ follows a sigmoidal dependence,
indicating successful learning. 
To further examine the behavior near $q=0$, the probability distribution
of $p^{*}_\mathrm{B}$ in the range of $-0.2<q<0.2$ is presented in
Figure~\ref{fig:committor}(b). 
For both LR and DNN, $p^{*}_\mathrm{B}$ exhibits a clear peak at 1/2. 
These results demonstrate that the RC $q$ predicted by the LR and DNN
models appropriately characterizes the isomerization reaction of alanine
dipeptide. 
The configurations near $q=0$, corresponding to $p_\mathrm{B}\approx 1/2$, can be
therefore regarded as the TS.

We quantified the contributions of grouped input variables to the RC $q$
predicted by the DNN using LIME and SHAP. 
Specifically, the dataset of 2,000 configurations was first partitioned
into three regions: 
$0 \le p_\mathrm{B}^* < 0.1$
(near state A), 
$0.49 < p_\mathrm{B}^{*} < 0.51$
 (near the TS), and 
$0.9 < p_\mathrm{B}^{*} \le 1$ (near state B). 
Within each region, 30 configurations were randomly selected and
analyzed using LIME and SHAP to quantify the contribution of each input
variable to the predicted RC.
Figure~\ref{fig:XAI} presents the absolute values of the average feature
contributions obtained from LIME and SHAP across the three regions. 
For comparison, the absolute values of the coefficients $(w_1,w_2,\cdots,w_{90})$
derived from the LR model are also shown in Figure~\ref{fig:XAI}. 
The feature contributions identified by LIME and SHAP exhibit trends
that closely match the coefficients obtained from linear regression. 
Table II of Ref.~\onlinecite{kikutsuji2022Explaining} lists the five
dihedral angles with the largest
absolute contribution values.

Indices 57 and 54, which rank highest in the LR model, correspond to 
$\sin \varphi$
(5–7–9–11) and $\sin \theta$
(6–7–8–9), respectively. 
This result indicates that the dihedral angle governing changes in
$\varphi$ is
$\theta$ rather than $\psi$.
As discussed above, comparison of Figures~\ref{fig:Ala2}(b) and
\ref{fig:Ala2}(c) shows that configurations with 
$p_\mathrm{B}^*\approx 1/2$
form a more distinct separatrix line on the $(\varphi, \theta)$
plane than on the $(\varphi, \psi)$
plane, consistent with this explanation.
Both LIME and SHAP further indicate that the dihedral angles $\varphi$
(Indices 11, 56, and 57) and 
$\theta$ (Indices 52 and 54) make substantial contributions to the
DNN-based prediction model. 
The results obtained from LIME and SHAP are largely consistent across
the three $p_\mathrm{B}^*$ regions, indicating agreement between LIME
and SHAP in the present analysis. 
Remarkably, both LIME and SHAP reveal 
that the most dominant contribution shifts from $\varphi$ to $\theta$
for configurations near the TS region ($0.49 < p_\mathrm{B}^{*} <
0.51$), in contrast to configurations
close to state A ($0 \le p_\mathrm{B}^* < 0.1$) or state B ($0.9 < p_\mathrm{B}^{*} \le 1$).
This distinction can be interpreted as follows:
near the TS, variations in $\theta$
induce a stronger influence than variations in $\varphi$, 
which accounts for the larger contribution to the conformation change.
Indeed, as shown in Figure~\ref{fig:Ala2}(c), the separatrix line defined by $p_\mathrm{B}^*=1/2$
is tilted with respect to the $\varphi$ axis.
This geometry indicates that variations in $\theta$
are essential for barrier crossing. 
Such specific TS features cannot be captured by the
global coefficients $(w_1,w_2,\cdots,w_{90})$ of the LR model.
Instead, they are revealed through local explanation models applied to
DNN predictions using XAI techniques such as LIME and SHAP.

\subsection{Hyperparameter tuning}
\label{sec:hyperparameter}

As demonstrated in Sec.\ref{sec:ala2}, DNN model can describe the RC quite effectively.
On the other hand, the hyperparameters, e.g. the number of hidden layers, nodes per hidden layer,
nd regularization penalty, need to be defined prior to optimizing the DNN model.
Choosing the hyperparameters can affect the performance of the DNN model and is a highly tedious task,
the adecuacy of the chosen hyperparameters remain ambiguous.
Neumann and Schwierz\cite{neumann2022Artificial} have applied the Keras Tuner random search
hyperparameter optimization to automatically determine the DNN model for the magnesium binding to RNA.
Nevertheless, how the choice of hyperparameters affects the quality and outcome of the DNN model remains unexplored.
Finding the appropriate set of hyperparameters thus remains highly tedious and nontrivial task.

The hyperparameter tuning framework using the Bayesian optimization method with a Gaussian process was applied to the isomerization reaction of alanine dipeptide in vacuum and in water.
The hyperparameter space of the DNN model for RCs in the two cases were optimized and explored.
The number of hidden layers ($N_\mathrm{layer}$), node per hidden layer ($\mathbf{N}_\mathrm{node}$), and $L_{2}$ regularization parameter for each hidden layer $\mathbf{\lambda}$ were chosen as the hyperparameters and determined automatically.
The data points with committor values and CVs were collected using the transition path sampling method.
The CV consists of 45 dihedral angles in cosine and sine forms following Sec.~\ref{sec:ala2} for the reaction in vacuum.
The electrostatic and vdW interactions from the waters to the atoms in alanine dipeptide were additionally adopted in the reaction in water.
The number of CVs were thus 90 and 134 for cases in vacuum and water, respectively.
The details of the system setup and sampling procedures are described in Ref. \onlinecite{kawashima2025Investigating}.
3714 and 4590 points were generated for the reactions in vacuum and in water, respectively.
States A and B were simply defined as $\varphi \le -30^\circ$ and $\varphi \ge 30^\circ$.

The hyperparameters were searched from 2 to 5 for $N_\mathrm{layer}$, from 100 to 5000 with 100 increments for $\mathbf{N}_\mathrm{node}$, and from 0.0001 to 0.1000 with 20 points equally spaced in a logarithmic scale for $\mathbf{\lambda}$, respectively.
Note that $\mathbf{N}_\mathrm{node}$ and $\mathbf{\lambda}$ were chosen separately for each hidden layer.
The initial values for these parameters were chosen randomly within the exploration range, and 10 models were constructed using different initial seeds for each system to see where the hyperparameters converge.
Bayesian optimization was performed for 150 trials, and DNN model was trained for a maximum of 1000 epochs using the cross-entropy function as the loss function for each hyperparameter set. Early stopping was applied when the value of the loss function did not improve for five consecutive steps. The data was divided into training and validation sets at a ratio of 8:2 and training, validation, and into training, validation, and test sets at a ratio of 5:1:4 during hyperparameter training and final DNN training, respectively.

The optimized hyperparameters for the reactions in vacuum and water are summarized in Table I and II of Ref. \onlinecite{kawashima2025Investigating}.
In both cases, the hyperparameters from different initial seeds ended up in different optimal parameter sets, thus did not converge to a unique minimum.
$N_\mathrm{node}$ most frequently converged to 5 and 3 in vacuum and in water, resepectively.
The number of nodes $\mathbf{N}_\mathrm{node}$ ranged from 100 to 5000 and differed between the hidden layers.
The regularization parameter $\lambda$ also converged to various values, and were generally larger in water than in vacuum.
The difference in $\lambda$ implies that the larger number of CV candiadates as input require larger penalty to suppress overfitting.
The root-mean-square errors (RMSE) between the predicted ($p_\mathrm{B}$) and reference ($p_\mathrm{B}^{*}$) committor values, shown in Fig. \ref{fig:rmse_train_test}, showed that RMSE in vacuum was smaller than those in water.
On the other hand, while the RMSEs for the training data showed some variations especially in the case in water, those for the test data were comparable between the RCs, e.g., the means were about 0.004 and 0.02 in vacuum and in water, respectively.
These results indicate that the RCs show similar performance despite the difference in the hyperparameters.
We also note that the RCs were found to be highly similar, e.g., the correlation coefficients between the coordinates calculated using the data points were mostly $>$ 0.99 (see Figs. S4 and S9 of Ref. \onlinecite{kawashima2025Investigating}).

The CVs contributing to the optimized RCs were also analyzed using LIME and SHAP.
Fig. \ref{fig:shap_vac_wat} summarizes the feature contributions of CVs in vacuum and water obtained with SHAP.
The features with notable contributions in vacuum were consistent with those obtained in Section~\ref{sec:ala2}.
Moreover, the RCs showed very similar trends as seen by the small variances.
In water, the electrostatic potential from water on H${}_{18}$, the hydrogen attached to the backbone nitrogen of C-terminal N-methylamide (Fig. \ref{fig:Ala2}a), were found to have large contribution as well as the torsion angles $\phi$ and $\psi$, represented by indices 54 and 61.
Interestingly, this resembles the solvent-derived electrostatic torque for a different path (C${}_{\mathrm{7eq}} \rightarrow \alpha_{\mathrm{R}}$ transition)\cite{ma2005Automatic}.
This suggests that the isomerization of a backbone of alanine dipeptide in water is induced by a water molecule interacting with the proton attached to the backbone, thereby triggering the twisting motion of the backbone.
Indeed, Fig. \ref{fig:correlation_3d}a shows that, depending on the position of the closest water to H${}_{18}$, the N-CA-C-N torsion angle is altered by the displacement of H${}_{18}$.
The correlation between CVs with indices 54, 61, and 125 and
$p_\mathrm{B}$, seen in the 3D scatter plot in
Fig.~\ref{fig:correlation_3d}b, also shows that CV${}_{125}$ (index 125)
is correlated with $p_\mathrm{B}$ and thus contributes to the RC.

\begin{figure}[t]
    \centering
    \includegraphics[width=0.7\linewidth]{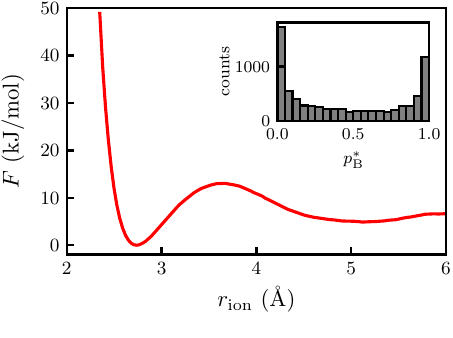}
    \caption{PMF $F(r_\mathrm{ion})$ as a function the interionic distance $r_\mathrm{ion}$.
The point with the minimum energy is set to 0 kJ/mol.
Inset: Distribution of Committor $p_\mathrm{B}^*$ evaluated for
 configurations sampled within the range $3.2~{\AA} < r_\mathrm{ion} <
 4.3$ {\AA}.
Reproduced from 
Okada \textit{et al.},
J. Chem. Phys. \textbf{164},
 094101 (2026)
with the permission of 
AIP
 Publishing LLC.~\protect\cite{okada2026Deep}
}
     \label{fig:ion_PMF}
\end{figure}

\begin{figure}[t]
    \centering
    \includegraphics[width=0.7\linewidth]{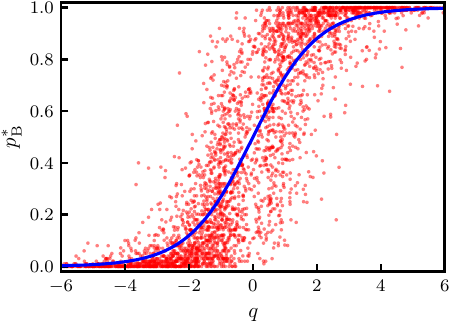}
    \caption{Committor $p_\mathrm{B}^*$ as a function of the
 RC $q$ predicted by DNN trained model using the test dataset (3,040 points).
The blue curve represents the sigmoidal function, $p_\mathrm{B}(q)=[1+\tanh(q)]/2$.
Reproduced from 
Okada \textit{et al.},
J. Chem. Phys. \textbf{164},
 094101 (2026)
with the permission of 
AIP
 Publishing LLC.~\protect\cite{okada2026Deep}
}
     \label{fig:ion_committor}
\end{figure}

\begin{figure*}[t]
    \centering
    \includegraphics[width=0.8\linewidth]{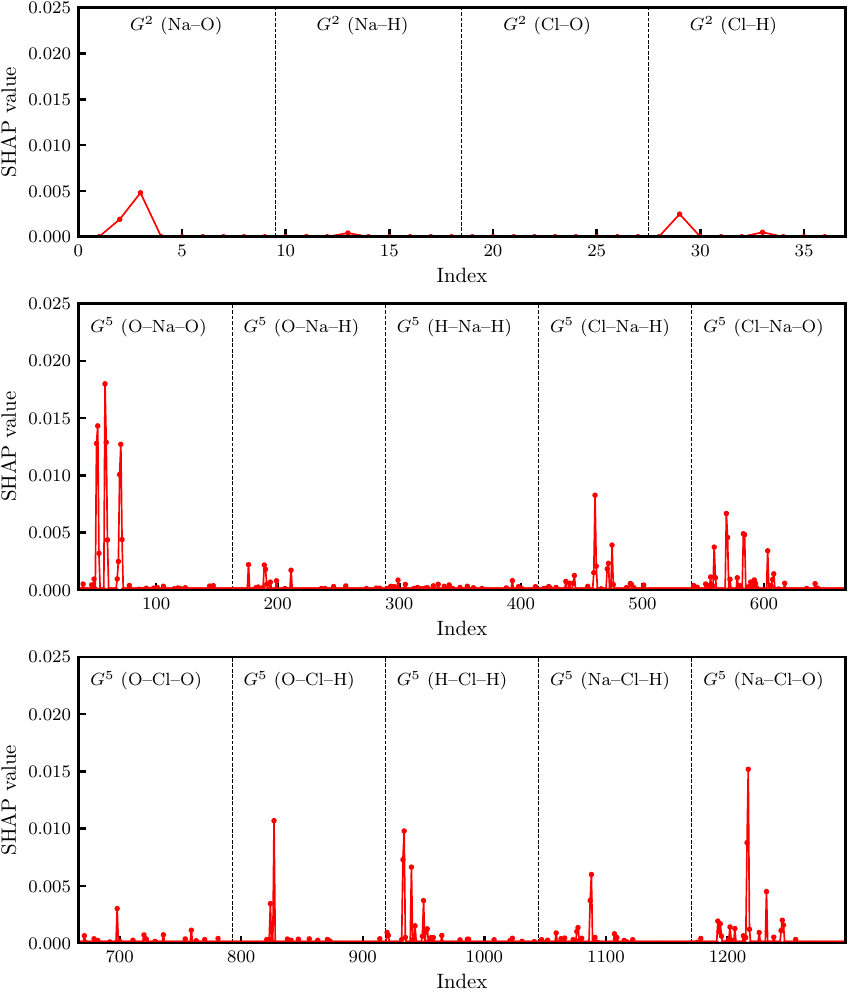}
    \caption{Index dependence of feature contribution of each CV evaluated by the absolute SHAP value.
Reproduced from 
Okada \textit{et al.},
J. Chem. Phys. \textbf{164},
 094101 (2026)
with the permission of 
AIP
 Publishing LLC.~\protect\cite{okada2026Deep}
}
     \label{fig:ion_SHAP}
\end{figure*}

\begin{figure}[t]
    \centering
    \includegraphics[width=\linewidth]{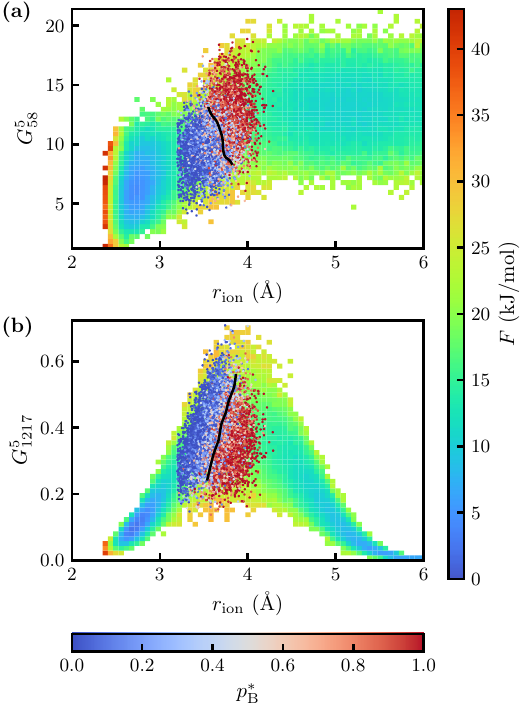}
    \caption{Two-dimensional PMF using the interionic
 distance $r_\mathrm{ion}$ and $G^5_{58}$ (a) and $G^5_{1217}$ (b).
The point with the minimum energy is set to 0 kJ/mol, and values are
 color-coded according to the color bar on the right.
Committor values $p_\mathrm{B}^*$ are colored according to the bottom color bar.
The black lines are drawn as follows.
The ranges of the $x$- and $y$-axis values were each divided into 200
 grid points (forming a 200 $\times$ 200 grid), 
and the average committor value $p_\mathrm{B}^*$ within each grid cell
 was computed. 
After applying cubic interpolation for smoothing, contour lines
 corresponding to $p_\mathrm{B}^*=0.5$ were plotted.
Reproduced from 
Okada \textit{et al.},
J. Chem. Phys. \textbf{164},
 094101 (2026)
with the permission of 
AIP
 Publishing LLC.~\protect\cite{okada2026Deep}
}
     \label{fig:ion_rion_G5}
\end{figure}

\begin{figure}[t]
    \centering
    \includegraphics[width=\linewidth]{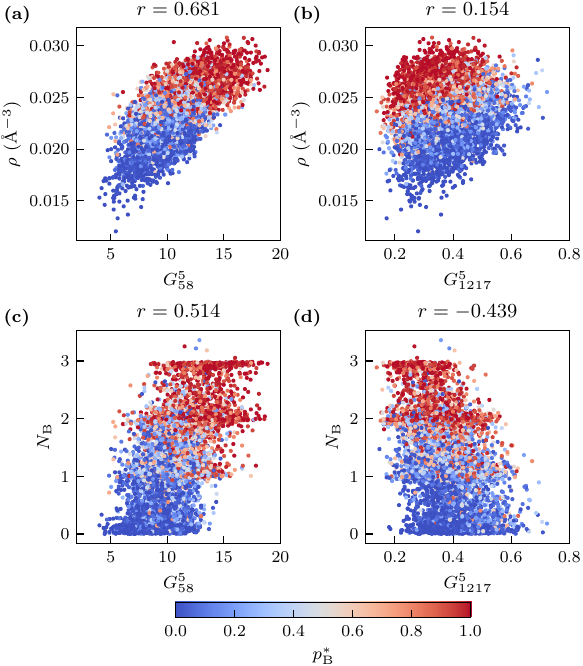}
    \caption{Distribution of committor $p_\mathrm{B}^*$ dataset in the
 two-dimensional surface 
using the
combinations $(G^5_{58}, \rho)$ (a), $(G^5_{1217}, \rho)$ (b),
$(G^5_{58}, N_\mathrm{B})$ (c), and $(G^5_{1217}, N_\mathrm{B})$ (d).
Committor values are colored according to the bottom color bar.
The $r$-value represents the correlation coefficient for each panel.
Reproduced from 
Okada \textit{et al.},
J. Chem. Phys. \textbf{164},
 094101 (2026)
with the permission of 
AIP
 Publishing LLC.~\protect\cite{okada2026Deep}
}
     \label{fig:ion_G5_rho_Nb}
\end{figure}

\subsection{Ion dissociation}
\label{sec:ion}

The association and dissociation of ion pairs in water are fundamental
processes in physical chemistry, yet their RCs are
complex, involving not only the interionic distance but also
solvent-mediated hydration structures.
This process may appear to be adequately described by the interionic
distance, $r_{\mathrm{ion}}$.~\cite{karim1986Dynamics, karim1986Ratea,
ciccotti1989Constrained, ciccotti1990Dynamics, guardia1991Potentiala,
rey1992Dynamical, smith1994Computer, brunig2022PairReaction}
However, $r_{\mathrm{ion}}$ alone does not fully capture the underlying
mechanism. 
Previous studies for NaCl ion pairs in water have shown that 
the committor distribution of configurations sampled near the saddle
point of the PMF defined $r_\mathrm{ion}$ is bimodal, with peaks near
$p_\mathrm{B}^*\approx 0$ and 1.~\cite{geissler1999Kinetic, ballard2012Mechanism}
This behavior indicates that configurations sharing the same interionic
distance $r_\mathrm{ion}$ can follow distinct transition pathways 
toward either association or dissociation, 
highlighting the essential role of 
the surrounding water structure in
determining the reaction.
Specifically, at the TS,
a bridging configuration is expected to form in which the O and H
atoms of a water molecule simultaneously coordinate to the Na and Cl
ions, respectively. 
Geissler \textit{et al.} proposed a mechanism in which a water molecule
initially coordinating the Na ion migrates toward the Cl-ion side,
thereby creating space for additional water molecules to enter.~\cite{geissler1999Kinetic}
Consequently, changes in the coordination number of water molecules near
the ions at the TS are thought to facilitate activation of
the dissociation process.

We applied an explainable deep learning framework to identify the RC for
the association and dissociation of a NaCl ion pair in TIP/4p-Ew water.
Detailed MD calculations are described in our prior study.~\cite{okada2026Deep}
The architecture of the DNN is the same as that used in Section~\ref{sec:ala2}.
A total of 7,600 configurations were extracted via umbrella sampling within the
interionic distance range, $3.2 \text{\AA} < r_{\mathrm{ion}} < 4.3 \text{\AA}$. 
This range was chosen to cover the saddle point located at
$r_{\mathrm{ion}} = 3.6$ {\AA} on the PMF defined by $r_{\mathrm{ion}}$,
as shown in Fig.~\ref{fig:ion_PMF}.
The associated state A is defined by $r_{\mathrm{ion}} < 3.2$ {\AA},
and the dissociated state B by $r_{\mathrm{ion}} >$ 4.3 {\AA}.
The committor value, $p_{\mathrm{B}}^{*}$ for each sampled configuration
was evaluated by assigning initial velocities randomly drawn from the
corresponding Maxwell--Boltzmann distribution of 300 K and performing
100 independent 1 ps MD simulations. 
The resulting distribution of $p_{\mathrm{B}}^{*}$ is shown in the inset
of Fig.~\ref{fig:ion_PMF}. 
The committor values are biased toward 0 and 1, with no pronounced peak
near 0.5, consistent with observations reported in previous
studies.~\cite{geissler1999Kinetic, ballard2012Mechanism}
In this study, two types of ACSFs describing the solvent environment
around the ions, $G^2$ and $G^5$, were employed as CVs for the neural
network inputs (see definitions of ACSFs in Section~\ref{sec:CV}).
For $G^2$, the cutoff radius was set to $R_\mathrm{c} = 10.0$ {\AA} with
$\eta = 2.0$ {\AA}$^{-2}$, while for $G^5$, $\eta = 1.2$ {\AA}$^{-2}$
was used. 
For $G^2$, four atomic combinations were considered: $(i-Z_1)$ = (Na-O),
(Na-H), (Cl-O), and (Cl-H). 
For $G_5$, the following 10 combinations were included: $(Z_1-i-Z_2)$ =
(O-Na-O), (O-Na-H), (H-Na-H), (Cl-Na-H), (Cl-Na-O), (O-Cl-O), (O-Cl-H),
(H-Cl-H), (Na-Cl-H), and (Na-Cl-O). 
By systematically varying the remaining parameters $R_\mathrm{s}$,
$\lambda$, and $\zeta$, we generated a total of 1,296 CVs
comprising 36 $G^2$ and 1,260 $G^5$ CVs. 
Detailed definitions of the $G^2$ and $G^5$ input features are provided
in Table S1 of Ref.~\onlinecite{okada2026Deep}.
The dataset was divided into training, validation, and test sets in a 5:1:4 ratio.
All input variables were standardized prior to neural network training.
Other computational details for DNN are described in our paper.~\cite{okada2026Deep}

Figure~\ref{fig:ion_committor} presents the learning results of the DNN
for the committor, $p_\mathrm{B}^{*}$. 
It is shown that the relationship between $p_\mathrm{B}^{*}$ and $q$
for the test dataset
has sufficiently converged to the sigmoidal function, $p_\mathrm{B}(q) = [1 + \tanh(q)]/2$.
However, the committor values near $q=0$ are
are widely distributed between 0 and 1 and do not exhibit a clear sharp peak at $p_\mathrm{B}^*=1/2$.
This behavior is likely due to the substantially larger number of input
variables compared to the alanine dipeptide case (see Section~\ref{sec:ala2}).
This limitation is expected to be improved in future studies
by not only improving the neural network training, including
hyperparameter tuning as discussed in Section~\ref{sec:hyperparameter},
but also through a more appropriate choice of descriptors for the local
solvation environment. 
In particular, the ACSFs used in the original study may not be
sufficient to fully capture the solvent
rearrangements relevant to ion dissociation, and more expressive
representations, such as graph-based representations,~\cite{kang2026Committors}
may therefore be worth exploring.

The contribution of each input variable to the RC $q$
predicted by the DNN was evaluated using SHAP. 
The absolute SHAP value for each index
was computed for 100 configurations randomly sampled from the
test dataset, and their average is shown in Fig.~\ref{fig:ion_SHAP}.
Tables I and II of Ref.~\onlinecite{okada2026Deep} show the top five contributing
factors and their corresponding absolute SHAP values for (O-Na-O) and
(Na-Cl-O) descriptors, respectively.

The $G^5$ at index 58 (denoted as $G^5_{58}$) shows the
largest contribution and corresponds to the O–Na–O combination, with
distance sensitivity $R_\mathrm{s} = 2.0$ {\AA}, angular sensitivity
$\lambda = -1$, and $\zeta = 1$. 
This indicates that water O atoms located within a spherical shell
of radius $R_\mathrm{s} = 2.0$ {\AA} centered on the Na ion make a
dominant contribution. 
Notably, $R_\mathrm{s} = 2.0$ {\AA} is nearly equal to the Lennard–Jones
diameter of Na ion. 
In contrast, the low angular resolution $\zeta = 1$ implies that angular
dependence plays a minor role in characterizing the distribution of
O atoms around Na ion.
The G5 at index 1217 (denoted as $G^5_{1217}$) 
corresponding to the (Na-Cl-O) combination, 
exhibits a second highest contribution following the RC and is characterized
by $R_\mathrm{s} = 4.0$ {\AA}, $\lambda = 1$, and $\zeta = 16$. 
At this high angular sensitivity, the angular component
$2^{1-\zeta}(1+\cos\theta_{ijk})^{\zeta}$ decreases from unity and
approaches zero near $\theta_{ijk} \approx 60^\circ$. 
This indicates that the dominant Na–Cl–O bond angles lie within this
range, suggesting
that water O atoms located in the overlapping region of the hydration
shells, namely, at a Na-Cl separation of $R_\mathrm{s} = 4.0$ {\AA} that
is slightly larger than the PMF saddle point distance $r_\mathrm{ion} =3.6$
{\AA}, are a primary contributing factor.

Based on these results, it is expected that $G^5_{58}$ or $G^5_{1217}$
will appropriately represent the RC together with the interionic distance
$r_\mathrm{ion}$.
Figure~\ref{fig:ion_rion_G5}(a) and (b) show the 
two-dimensional PMFs constructed using ($r_\mathrm{ion}$, $G^5_{58}$)
and ($r_\mathrm{ion}$, $G^5_{1217}$), respectively.
The 
distribution of
committors $p_\mathrm{B}^*$ are also shown on the corresponding $(r_\mathrm{ion}, G^5_{58})$ and
$(r_\mathrm{ion}, G^5_{1217})$ planes.
In these figures, the committor $p_{\mathrm{B}}^{*}$ varies continuously from the
associated state A $(p_{\mathrm{B}}^{*} \approx 0)$ to the dissociated
state B $(p_{\mathrm{B}}^{*} \approx 1)$, with the TS
$(p_{\mathrm{B}}^{*} = 0.5)$ forming a well defined separatrix line. 
Figure~\ref{fig:ion_rion_G5}(a) shows that, when the interionic distance
$r_\mathrm{ion}$ is fixed at the saddle point of the PMF,
$r_\mathrm{ion} = 3.6$ {\AA}, the committor increases with increasing $G^5_{58}$,
indicating that the arrangement of water O atoms surrounding the
Na ion plays a key role in ion pair dissociation. 
In contrast, Fig.~\ref{fig:ion_rion_G5}(b) shows that, 
starting from the associated state ($p_\mathrm{B}\approx 0$),
$G^5_{1217}$ initially increases together with 
$r_\mathrm{ion}$ as the system approaches the TS ($p_\mathrm{B}=0.5$). 
Upon crossing the separatrix line, however, $G^5_{1217}$ decreases as 
the system proceeds toward the dissociated state ($p_\mathrm{B}\approx 1$).
This behavior implies that dissociation is
associated with a reduction in the population of water O atoms
simultaneously belonging to the Na and Cl hydration shells. 
Thus, the two ACSFs, $G^5_{58}$ and $G^5_{1217}$, identified through the
SHAP analysis highlight
the central role of solvent environments in the association
and dissociation of NaCl ion pairs in water.

Finally, we conducted a comparative analysis of the correlation between
$G^5_{58}$ and $G^5_{1217}$ and 
$\rho$ and $N_\mathrm{B}$, identified as CVs
representing water bridging structures (see definitions of $\rho$ and
$N_\mathrm{B}$ in Section~\ref{sec:CV}).~\cite{mullen2014Transmission}
Figure~\ref{fig:ion_G5_rho_Nb} shows the distribution of the committor
$p_{\mathrm{B}}^{*}$ on
two-dimensional planes defined by the variable combinations $(G^5_{58},
\rho)$ (a), $(G^5_{1217}, \rho)$ (b), $(G^5_{58}, N_\mathrm{B})$ (c),
and $(G^5_{1217}, N_\mathrm{B})$ (d). 
The correlation coefficient, $r$, for each variable pair is also
reported in Fig.~\ref{fig:ion_G5_rho_Nb}. 
The interionic water density $\rho$ exhibits a positive correlation with
$G^5_{58}$ and a weak negative correlation with $G^5_{1217})$. 
Similarly, the number of bridging water molecules $N_\mathrm{B}$ shows a
positive correlation with $G^5_{58}$, while its correlation with
$G^5_{1217}$ is slightly stronger than that observed for $\rho$. 
Thus, increases in $\rho$ and $N_\mathrm{B}$, which correlate positively
with $G^5_{58}$ and negatively with $G^5_{1217}$, are associated with
dissociation from the associated state, consistent with the separatrix
line observed in Fig.~\ref{fig:ion_rion_G5}. 
Moreover, the committor distributions in each panel of
Fig.~\ref{fig:ion_rion_G5} display a clear separation between associated
and dissociated states. 
Overall, the two CVs, $\rho$ and $N_\mathrm{B}$, are confirmed as key
contributors to the RC through their correspondence with the two ACSF
descriptors $G^5_{58}$ and $G^5_{1217}$.

\section{Summary and outlook}

In this review, we have presented an explainable deep learning framework
for identifying RCs in complex molecular systems based on 
committor $p_\mathrm{B}^*$ sampling.
By combining DNNs with model-agnostic interpretation
techniques such as LIME and SHAP, this framework enables the
identification of CVs that contribute significantly to the RC while
maintaining interpretability.
Applications to representative systems, including the isomerization of
alanine dipeptide and the association and dissociation of NaCl ion pairs
in water, demonstrate that the framework can successfully extract
physically meaningful features governing their transition processes.

Furthermore, DNN models with diverse architectures, differing in the
number of nodes, layers, and regularization parameters, 
describe the RC
with comparable accuracy.
The CVs identified through LIME and SHAP are highly
consistent across different models, indicating that the hypaerparameter
space of the neural network is multimodal.
This suggests that, although multiple model structures may achieve
similar predictive performance, the underlying physical features that
govern the transition mechanism remain robust.

These results highlight the potential of the explainable deep learning
framework to systematically identify relevant CVs from high-dimensional
configurational spaces.
Compared with conventional approaches for describing PMFs, which often
rely 
heavily on physical intuition and trial and error, the present framework
provides a more data-driven strategy for constructing PMFs and
interpreting the mechanism of transition pathways.
Nevertheless, the reliability of the identified RCs depends on the
selection of candidate CVs and the performance of the neural network
model, emphasizing the importance of careful feature design 
and appropriate hyperparameter optimization.

The examples discussed here also reflect a methodological
progression that has been common in committor-related studies. 
Alanine dipeptide has long served as a benchmark system for testing RC and
committor learning methodologies in a well-controlled conformational
transition problem. 
From this starting point, ion dissociation in solution provides a
natural extension toward processes involving solvent degrees of freedom
and solvation rearrangements. 
The present examples were therefore selected not simply as two
independent model systems, but as representative steps in the
development of committor learning approaches from a benchmark
conformational 
transition to solvent-involving molecular processes. 
At the same time, these examples remain illustrative rather than
exhaustive, and more complex conformational sampling problems will
likely require further validation of both the learned committor models
and their interpretation.

Future developments are expected to further enhance the
applicability of this
framework to increasingly complex systems. 
In particular, improvements in the definition of CVs, integration with advanced
sampling techniques, and the incorporation of more advanced machine
learning architectures, such as graph neural
networks,~\cite{zhang2024DescriptorFree, pengmei2025Using,
kang2026Committors, contrerasarredondo2026Learning} will enable the
analysis of larger and more
complex molecular environments. 
A particularly promising direction is the development of
enhanced sampling methods that utilize the committor together with
variational principles.
For example, the application of the Kolmogorov variational principle
makes it possible to sample the committor as RC
under boundary conditions in which it is fixed to 0 in state A and 1 in
state B.~\cite{kang2026Committors}
Another variational principle is based on representing the committor
function in terms of its time-correlation
function.~\cite{contrerasarredondo2026Learning}
In addition, the development of methods that combine interpretability
with automated feature generation may reduce the dependence on
predefined CVs. 
These advances will contribute to a deeper understanding of rare-event
mechanisms in complex molecular systems and will expand the role of
explainable deep learning in theoretical and computational chemistry.

\tocless{\section*{Acknowledgments}}
The authors acknowledge Yusuke Mori, Takuma Kikutsuji, Kazushi Okada,
Kyohei Kawashima, Takumi Sato, 
 and Kenji Okada for 
fruitful collaborations.
This work was supported by 
JSPS KAKENHI Grant-in-Aid 
(Grant Nos.~\mbox{JP23K23858}, \mbox{JP25H02299}, \mbox{JP23K23303},
 \mbox{JP23KK0254}, \mbox{JP24K21756},
\mbox{JP25H02464}, \mbox{JP25K02235}, 
\mbox{JP25K00968}, \mbox{JP24H01719}, 
 \mbox{JP22K03550}, and \mbox{JP23H02622}) and Shiraishi Science
 Promotion Association.
We acknowledge support from
the Fugaku Supercomputing Project (Nos.~\mbox{JPMXP1020230325} and \mbox{JPMXP1020230327}) and 
the Data-Driven Material Research Project (No.~\mbox{JPMXP1122714694})
from the
Ministry of Education, Culture, Sports, Science, and Technology
and by
Maruho Collaborative Project for Theoretical Pharmaceutics.
The numerical calculations were performed at Research Center of
Computational Science, Okazaki Research Facilities, National Institutes
of Natural Sciences (Projects: 25-IMS-C052, 25-IMS-C105, 25-IMS-C227,
26-IMS-C019, 26-IMS-C051, and 26-IMS-239).

\tocless{\subsection*{AUTHOR DECLARATIONS}}

\tocless{\subsection*{Conflict of Interest}}
The authors have no conflicts to disclose.

\tocless{\subsection*{Data availability statement}}
No new data were generated or analyzed in this study; therefore, data sharing is not applicable.

%

\end{document}